\documentclass[twocolumn,amsmath,amssymb,showpacs]{revtex4}
\usepackage{graphicx}

\def\d{{\partial}}
\def\s{{\sigma}}
\def\e{{\epsilon}}
\def\k{{ {\bf k} }}

\def\w{{\omega}}
\def\a{{\alpha}}
\def\b{{\beta}}

\def\g{{\gamma}}
\def\l{{\lambda}}

\begin{document}

\def\runtitle{
Intrinsic anomalous Hall effect in ferromagnetic metals \\
studied by the multi-\textit{d}-orbital tight-binding model
}
\def\runauthor{
Hiroshi {\sc Kontani}$^1$, Takuro {\sc Tanaka}$^1$ and Kosaku {\sc Yamada}$^2$
}

\title{
Intrinsic anomalous Hall effect in ferromagnetic metals \\
studied by the multi-\textit{d}-orbital tight-binding model
}

\author{
Hiroshi {\sc Kontani}$^1$, Takuro {\sc Tanaka}$^1$ and Kosaku {\sc Yamada}$^2$
}

\address{
$^1$Department of Physics, Nagoya University,
Furo-cho, Nagoya 464-8602, Japan. \\
$^2$Engineering, Ritsumeikan University,
1-1-1 Noji Higashi, Kusatsu, Shiga 525-8577, Japan.
}

\date{\today}

\begin{abstract}
To elucidate the origin of anomalous Hall effect (AHE) 
in ferromagnetic transition metals, 
we study the intrinsic AHE based on a 
multi-orbital ($(d_{xz},d_{yz})$) tight-binding model.
We find that a large anomalous velocity comes from the off-diagonal
(inter-orbital) hopping.
By this reason, the present model shows a large intrinsic anomalous 
Hall conductivity (AHC) which is compatible with typical experimental 
values in ferromagnets [$10^2\sim10^3\Omega^{-1}{\rm cm}^{-1}$], 
without necessity to assume a special band structure at the Fermi level.
In good metals where $\rho$ is small, 
the intrinsic AHC is constant (dissipation-less) 
as found by Karplus and Luttinger.
In bad metals, however, we find that the AHC is proportional 
to $\rho^{-2}$ when $\hbar/2\tau$ is larger than the minimum
band-splitting measured from the Fermi level, $\Delta$.
This crossover behavior of the intrinsic AHE,
which was first derived in J. Phys. Soc. Jpn. {\bf 63} (1994) 2627,
is recently observed in various ferromagnetic metals universally
by A. Asamitsu et al.
We also stress that the present $(d_{xz},d_{yz})$-tight binding model 
shows a huge spin Hall effect in a paramagnetic state. 
\end{abstract}

\sloppy

\pacs{72.10.-d, 72.80.Ga, 72.25.Ba}

\maketitle


\section{Introduction}
 \label{Intro}
\subsection{Motivation and Purpose of the Study}

For a long time, the anomalous Hall effect (AHE) attracts increasing
attraction from both theoretical and experimental viewpoint.
In general, the Hall resistivity is given by
$\rho_H = R_H^0B + 4\pi R_H^{\rm a}M$,
where $B$ is the magnetic field and $R_H^0$ is the ordinary
Hall coefficient due to the Lorentz force.
$R_H^{\rm a}$ is the anomalous Hall coefficient
in the presence of magnetization $M$.
In ferromagnets, $R_H^{\rm a}$ is usually at least
one order of magnitude larger than $R_H^0$.
In paramagnetic heavy fermion (HF) systems,
$\rho_H/B$ takes a huge value due to the AHE
since the uniform susceptibility $M/B=\chi$ in HF 
is about $10^2\sim10^3$ times larger than that in 
usual metals owing to the strong Coulomb interaction.

Recently, the AHE in transition metal ferromagnets
has been intensively studied experimentally.
For example, the AHE in ferromagnetic Pyrochlore oxides shows
interesting behaviors
 \cite{Yoshii,Yasui,Taguchi1,Taguchi2}.
The AHE is also observed in ferromagnetic spinels \cite{Ong} 
and in Ru, Ti oxides \cite{Reiner,Izumi}.
Several multilayer systems \cite{HSato2,Canedy} also show distinct AHE.
The AHE in spin glass systems had been discussed in refs.
 \cite{Tatara,Kawamura}.

The theoretical study of AHE was initiated by Karplus and 
Luttinger (KL) \cite{KL} in 1954.
They found that the anomalous Hall conductivity (AHC)
$\s_{xy}^{\rm a} \ (=R_H^{\rm a}M/\rho^2)$ is finite 
and dissipation-less (i.e., $\s_{xy}^{\rm a}$ is independent of 
resistivity $\rho$) when $M\ne 0$.
This KL-term is called the ``intrinsic AHE'' because it exists 
independent of impurities.
In 1958, Smit presented a mechanism of ``extrinsic AHE'' \cite{Smit}:  
He found that spin polarized electrons are scattered asymmetrically 
around an impurity in the presence of spin-orbit coupling.
The AHC due to this skew-scattering mechanism is 
linearly proportional to $\rho$ if elastic scattering is dominant.
The above two works were reproduced and refined
in a unified way by Luttinger \cite{KL2} by using the transport 
theory of Kohn and Luttinger,
and his work were also reproduced by linear-response theory
using diagrammatic technique \cite{Fukuyama,Sinitsyn2}.
In 1970, Berger proposed another mechanism of extrinsic AHE,
the side jump due to impurities \cite{Berger}.
This mechanism gives the AHC in proportion to $\rho^2$.
Note that the extrinsic AHE vanishes in a clean system without impurity.

For a long time, the AHE had been generally regarded as 
an extrinsic effect, and the intrinsic AHE proposed by KL
had been underestimated.
This would be because the theoretical model assumed
in ref. \cite{KL2} was too oversimplified.
Moreover, ref. \cite{KL2} could not offer a specific 
expression for $\s_{xy}^a$.
Recently, however, various experiments suggest that
the intrinsic AHE $\s_{xy}^{\rm a} \propto (\rho)^0$
is dominant in many transition metal ferromagnets
 \cite{Ong,Asamitsu} even in good metals with $\rho\sim1\mu\Omega{\rm cm}$:
It is surprising because the extrinsic AHC $(\propto \rho^{-1})$
should be dominant when $\rho$ is small,
if inelastic scattering is negligibly small.
We here remind readers that the extrinsic AHC is very sensitive to
the statistical property of impurity potential $V$;
For instance, it vanishes when the average value of $V^3$ is zero
 \cite{Bruno}.

After Luttinger's work \cite{KL2}, theory of intrinsic AHE
has been developed by many authors.
In general, $\s_{xy}$ is composed of the ``Fermi surface term''  
and the ``Fermi sea term''; the latter comes from quasiparticles 
inside of the Fermi sea, and it could exist even in insulators
\cite{Streda}.
(Note that the conductivity $\s_{xx}$ is composed of only 
the Fermi surface term.)
The term of the AHC derived by KL is a part of the ``Fermi sea term''.
Recently, M. Onoda and Nagaosa \cite{MOnoda} and 
Sundaram and Niu \cite{Niu} found that KL's AHC 
is expressed in terms of the ``Berry curvature''
 \cite{Niu,MOnoda}.
(Note that Luttinger showed in his model that there is a term 
which almost cancels the Berry curvature term \cite{KL2}.)
On the other hand,
Kontani and Yamada derived the intrinsic AHE due to the 
``Fermi surface term'' based on the linear-response theory
for the first time \cite{Kontani94,Kontani97}.
They studied AHC in an orbitally degenerate periodic Anderson model,
and succeeded in explaining the AHE in HF systems;
$\s_{xy}^a \propto \chi$ below the coherent temperature $T_0$, 
whereas $\s_{xy}^a \propto \chi\rho^{-2}$ above $T_0$
 \cite{Onuki1,Onuki2}.
This results explain the experimental fact that the Hall 
coefficients in HF systems are proportional to $\rho^2$ below $T_0$.
Miyazawa et al. proved that the intrinsic AHE also occurs
in $d$-$p$ models with orbital degree of freedoms
 \cite{Miyazawa}.
Later, AHC's for Fe and SrRuO$_3$
were calculated based on the LDA band calculations
 \cite{Fe,Fang}.

We still have to deepen the understanding of
the mechanism of AHE to explain experimental results
in transition ferromagnetic metals.
For example, a recent experiment by Asamitsu et al. has revealed that
the intrinsic AHC $(\s_{xy}^{\rm a}\sim 10^3 \Omega^{-1}{\rm cm}^{-1})$
is observed in many ferromagnets for 
$\rho=1 \sim 100\mu\Omega{\rm cm}$,
whereas $\s_{xy}^{\rm a}$ starts to decrease in proportion to 
$\rho^{-n}$ and $n\sim2$ in bad metals where $\rho\gg 100\mu\Omega{\rm cm}$.
This drastic crossover is reminiscent of the AHE in 
heavy fermion systems discussed in ref. \cite{Kontani94}.
Therefore, a detailed study based on an appropriate model
for transition metals is highly required.

In the present paper, we study the intrinsic AHC in a 
tight-binding model with $(d_{xz},d_{yz})$-orbitals ($e_g'$-orbitals),
which originate from the $t_{2g}$-orbitals $(d_{xz},d_{yz},d_{xy})$
in the tetragonal crystalline field.
The band structure of this model corresponds to  
$\a$ and $\b$ bands of Sr$_2$RuO$_4$ \cite{Miyazawa,Sigrist,Nomura,Yanase}.
We derive a general expression for the AHC valid for any
damping rate $\hbar/2\tau$, 
which enables us to study the AHC in bad metals.
We show that the AHC is mainly given by the Fermi surface term 
for a wide range of $\hbar/2\tau$
since the KL's term is canceled by another Fermi sea term
in a metallic state.
We also find that the anomalous velocity due to $d(xz)$-$d(yz)$
hopping gives rise to a large AHC comparable to experimental 
values (about $e^2/ha$; $a$ being the lattice constant),
which will be a main origin of a huge AHC in transition
metal ferromagnets.
In good metals, the intrinsic AHC is independent of $\tau$
as is well known.
However, it becomes proportional to $\rho^{-2}$ in bad metals 
where $\hbar/2\tau$ is larger than the band-splitting 
around the Fermi level, $\Delta$.
This crossover behavior of the AHC can explain a recent observation 
of the AHE by A. Asamitsu et al.

\subsection{Origin of Intrinsic and Extrinsic AHE}
Here, we shortly explain the mechanisms of both the 
intrinsic AHE and the extrinsic AHE in more detail:
The intrinsic AHE (both Fermi surface and Fermi sea terms) 
originates from the interband transition of quasiparticles 
due to off-diagonal terms of the velocity ${\hat v}_\mu$ and 
the orbital angular momentum ${\hat l}$.
(The diagrammatic expression for $\s_{xy}^a$ is given in fig. 
\ref{fig:diagram}.)
The off-diagonal velocity contains the ``anomalous velocity'' 
(e.g. $v_x^{\rm a} \propto k_y$), which gives rise to the AHE.
Since it is a purely quantum effect, it is difficult to find a 
simple classical analogue except for Rashba-type 2D electron gas model
 \cite{Sinitsyn1}.
This might be a reason why the intrinsic AHE had been discounted
for a long time.
In the present model, a large anomalous velocity naturally
comes from the inter-orbital ($d_{xz}$-$d_{yz}$) hopping.
By this reason, atomic $d$-orbitals are necessary to reproduce
a huge AHC in real ferromagnets.

On the other hand, the extrinsic AHE (the skew-scattering mechanism 
and the side-jump mechanism) happens even in a single-band model,
although it is enhanced due to a multiband effect
 \cite{Smit,Berger}.
Extrinsic AHCs due to skew-scattering mechanism and 
the side-jump mechanism
are proportional to $\rho^{-1}$ and $(\rho)^0$, respectively.
The former is caused by spin-dependent asymmetric scattering 
around the impurity in the presence of spin-orbit coupling,
and the latter is caused by lateral displacement of the 
wave-packet during the scattering
 \cite{Smit,Berger,Bruno,Kondo}.
Diagrammatically, both skew-scattering and side-jump are
expressed by {\it current vertex corrections} due to impurity potential,
as shown in Ref. \cite{Bruno}.
In general, detailed knowledge on the impurity potential
is needed for a quantitative study of the extrinsic AHE.
In the present paper, we do not study the extrinsic AHE
since recent several experiments suggest that the 
skew scattering term is small even in good metals
 \cite{Ong,Asamitsu}.

For a long time, it was believed that the AHE is independent of 
the {\it self-energy correction} due to impurity potential.
However, we show that the intrinsic AHC is reduced by 
the self-energy correction in bad metal.
By this reason, change of the scaling law in the AHC
at $\rho \sim 100\mu\Omega$cm reported by Asamitsu el at.
 \cite{Asamitsu}.
is well explained in terms of the intrinsic AHE.

\section{Model and Hamiltonian}
  \label{Model}
In the present paper,
we study a square lattice tight-binding model with $d_{xz}$ and $d_{yz}$
orbitals, which is a simplified version of the $(d_{xz},d_{yz})$-$p_z$ model.
Miyazawa et al. \cite{Miyazawa} showed that the AHE exists
in the latter model.
However, they could not derive an explicit expression for AHC.
Here, we derive explicit expressions for both the ``Fermi surface
term'' and the ``Fermi sea term'' based on the present simplified model.

The $ls$-coupling term is indispensable for the AHE in ferromagnets:
In a ferromagnetic metal with ${\bf M} \parallel {\hat z}$, 
the $ls$-coupling term $\l{\bf l} \cdot {\bf s}$
is approximately proportional to $-\l M {\hat l}_z$, 
where $\l$ is the coupling constant.
The AHE is caused by the inter-orbital transition
of quasiparticles due to the off-diagonal elements of ${\hat l}_z$
 \cite{KL}.
In a paramagnetic metal under $B_z$, on the other hand, 
Zeeman term for the angular momentum, $\mu_{\rm B} B_z {\hat l}_z$,
gives rise to the AHE
 \cite{Kontani94}.
Therefore, the AHE in paramagnetic metals and that
in ferromagnetic ones are caused by the same transport mechanism,
although the origins of magnetizations are different.

We consider that the present model with $d_{xz}$- and $d_{yz}$-orbitals 
describes a major part of the AHE in transition metal ferromagnets:
In cubic or tetragonal crystals,
$\langle \a|{\hat l}_z|\b \rangle$ is nonzero
only when $(\a,\b)=(xz,yz)$ $[l_z=\pm 1]$ and 
$(\a,\b)=(xy,x^2-y^2)$ $[l_z=\pm 2]$.
Because levels of $d_{xz}$ and $d_{yz}$ orbitals are degenerate,
the band structures composed of these orbitals are energetically close.
This fact will be favorable to the interband hopping of quasiparticles 
between $d_{xz}$ and $d_{yz}$ orbitals, which is indispensable for the AHE.
On the other hand,
energy splitting between $d_{x^2-y^2}$-orbital and 
$d_{xy}$-orbital is of the order of 1eV in square or tetragonal crystals.
Therefore, the interband hopping of quasiparticles 
between $d_{x^2-y^2}$- and $d_{xy}$-orbitals would be difficult.
As a result, the main contribution to the AHE in transition
metal oxides will come from the inter-orbital transition 
between $d_{xz}$- and $d_{yz}$-orbitals.
This is the reason why we study the $(d_{xz},d_{yz})$-orbital 
tight-binding model.

\begin{figure}[htbp]
\includegraphics[width=.8\linewidth]{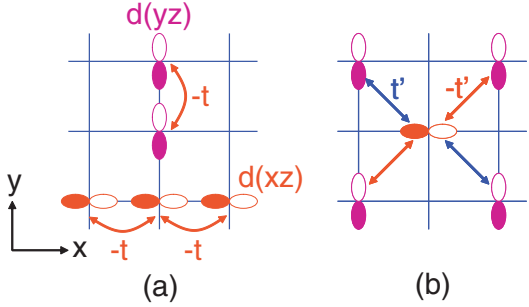}
\caption{(color online) (a) Hopping integrals between the same orbitals.
(b) Hopping integrals between the different orbitals,
which is given by the next nearest neighbor hopping.
The sign of the hopping integral changes by $\pi/2$ rotation.
This fact gives rise to the anomalous velocity.
}
  \label{fig:model}
\end{figure}

Here, we represent the creation
operator of an electron on $xz$ ($yz$) orbital as
${c_\k^{x}}^\dagger$ (${c_\k^{y}}^\dagger$).
The Hamiltonian without $ls$-coupling is given by
$H^0= \sum_\k {\hat c}_\k^\dagger {\hat h}_\k^0 {\hat c}_\k$,
where
\begin{eqnarray}
{\hat h}_\k^0&=&\left(
\begin{array}{cc}
\xi_\k^{x}& \xi_\k^{xy} \\
\xi_\k^{xy} & \xi_\k^y 
\end{array} 
\right) ,
 \label{eqn:H0}
\end{eqnarray}
and ${\hat c}_\k^\dagger= ({c_\k^{x}}^\dagger,{c_\k^{y}}^\dagger)$.
$\xi_{\k}^{x}=-2t\cos k_x$, 
$\xi_{\k}^{y}=-2t\cos k_y$ and
$\xi_{\k}^{xy}= 4t'\sin k_x \sin k_y$.
Here, $-t$ and $\pm t'$ are the hopping integrals
between nearest-neighbors and next-nearest-neighbors, respectively
 \cite{Nomura}.
They are shown in fig. \ref{fig:model}.

Then, the velocity matrix is given by
${\hat v}_\mu = \d {\hat h}_\k^0 / \d k_\mu$,
where $\mu=x,y$.
They are given by
\begin{eqnarray}
{\hat v}_x 
&=&\left(
\begin{array}{cc}
2t\sin k_x & 4t'\cos k_x \sin k_y \\
4t'\cos k_x \sin k_y & 0 
\end{array} 
\right) ,
 \label{eqn:vx} \\
{\hat v}_y
&=&\left(
\begin{array}{cc}
0 & 4t'\sin k_x \cos k_y \\
4t'\sin k_x \cos k_y & 2t\sin k_y 
\end{array} 
\right) .
 \label{eqn:vy}
\end{eqnarray}
We should stress that the off-diagonal elements of ${\hat v}_x$,
$v_x^{xy}=v_x^{yx}$, is an odd-function of $k_y$.
In the same way, $v_y^{xy}=v_y^{yx}$ is an odd-function of $k_x$.
They are called ``anomalous velocity''.
In later sections, we will show that $\s_{xy}^{\rm a}$ is proportional
to $\langle v_x^{xy}v_y^{yy}\rangle$, which can remain finite
after the $\k$-summation due to the anomalous velocity.
Therefore, a sizable AHC is caused by $v_x^{xy}$ and $v_y^{xy}$.
Atomic $d$-orbitals give rise to the huge AHC in 
transition metal ferromagnets.

To realize the AHE in ferromagnets, the atomic $ls$-coupling is 
also necessary.
It is given by 
$H^\l= \sum_\k {\hat c}_\k^\dagger {\hat h}^\l {\hat c}_\k$.
Because $|xz\rangle= -(|l_z=+1\rangle - |l_z=-1\rangle )/\sqrt{2}$
and $|yz\rangle= i(|l_z=+1\rangle + |l_z=-1\rangle )/\sqrt{2}i$,
${\hat h}^{\l}$ in the present basis is given by
\begin{eqnarray}
 {\hat h}^{\l}&=& {\rm sgn}(s_z) \l {\hat \tau}_y ,
\end{eqnarray}
where $\l$ is the coupling constant and ${\hat \tau}_y$
is the Pauli matrix for the orbital space.
Note that ${\hat l}_y={\hat l}_z=0$ in the present basis.
In Fe, $\l=70$meV
 \cite{Fe}.
Hereafter, we put $\mu_{\rm B}=1$ for the simplicity of calculation.

\begin{figure}[htbp]
\includegraphics[width=.8\linewidth]{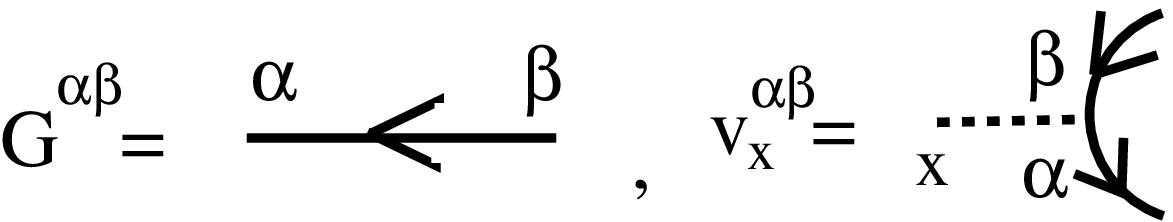}
\caption{Diagrammatic expression for the Green function $G_{\a\b}$ 
and the velocity $v_x^{\a\b}$.
}
  \label{fig:AHE-Green}
\end{figure}
\begin{figure}[htbp]
\includegraphics[width=.7\linewidth]{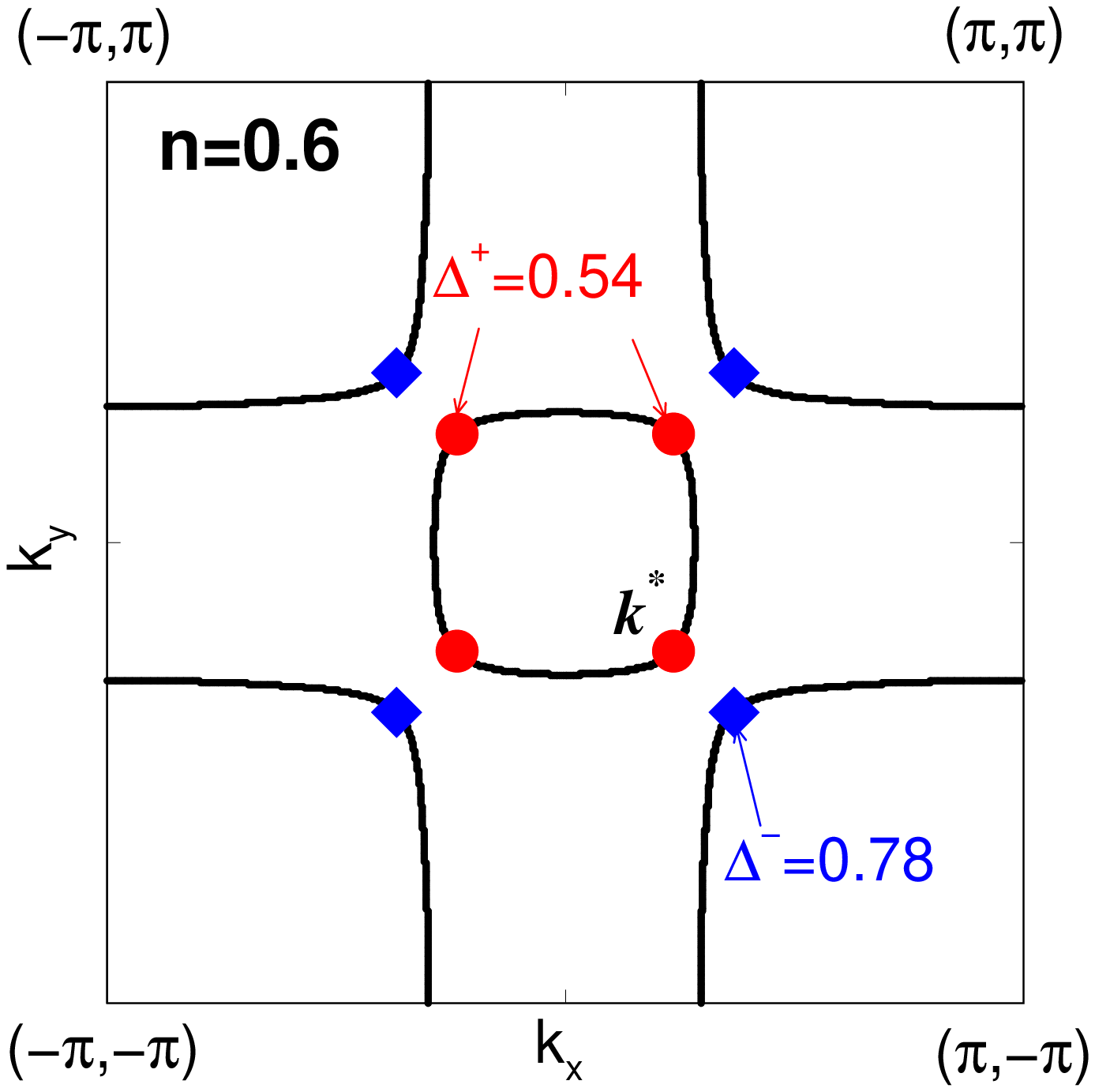}
\includegraphics[width=.7\linewidth]{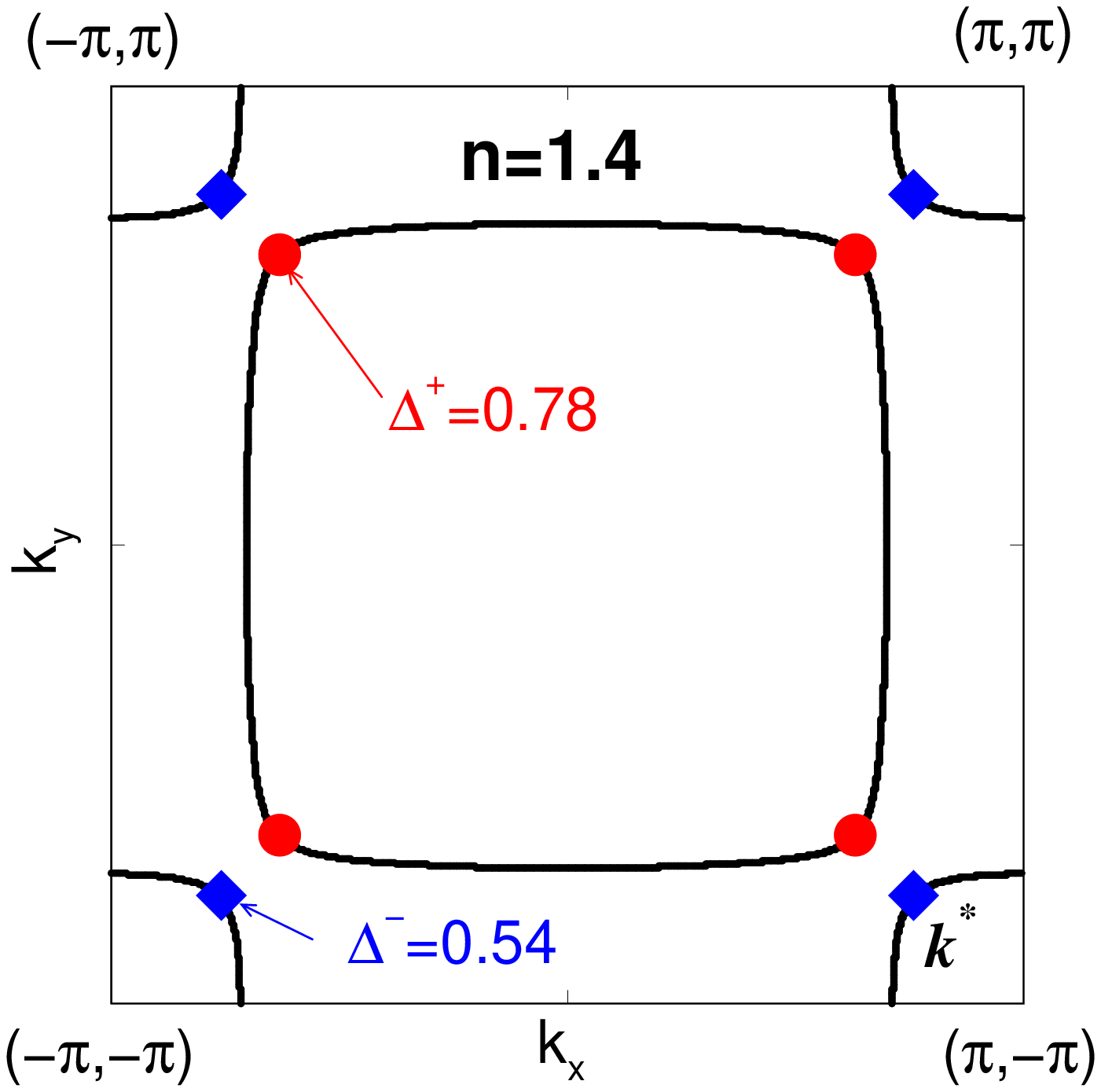}
\caption{(color online) Fermi surfaces for (a) $n=0.6$ and (a) $n=1.4$.
Here we put $(t,t')=(1,0.1)$, which corresponds to Sr$_2$RuO$_4$.
}
  \label{fig:FS}
\end{figure}

The Green function in the presence of atomic $ls$-coupling is given by
${\hat G}_\k(\w)= (\w+\mu-{\hat h}_\k^0-{\hat h}^{\l})^{-1}$,
which is given by
\begin{eqnarray}
& &\left(
\begin{array}{cc}
G_{xx} & G_{xy} \\
G_{yx} & G_{yy} 
\end{array} 
\right) \nonumber \\
& &\ \ \ \ \ 
= \frac1{d(\w)}
\left(
\begin{array}{cc}
\w+\mu-\xi_\k^y& \a_\k \\
\a_\k^* & \w+\mu-\xi_\k^x
\end{array} 
\right) ,
 \label{eqn:Green}
\end{eqnarray}
where $\a_\k= \xi_\k^{xy} + i\l {\rm sgn}(s_z)$ and 
$d(\w)=(\w+\mu-\xi_\k^x)(\w+\mu-\xi_\k^y)-|\a_\k|^2$,
which is expressed as
\begin{eqnarray}
d(\w)&=&(\w+\mu-E_\k^+)(\w+\mu-E_\k^-),
 \\
E_\k^\pm&=&\frac12\left( \xi_\k^x+\xi_\k^y 
 \pm \sqrt{(\xi_\k^x-\xi_\k^y)^2+4|\a_\k|^2} \right) ,
\end{eqnarray}
where $E_\k^\pm$ represents the quasiparticle dispersion.
Figure \ref{fig:FS} shows the Fermi surfaces for $(t,t')=(1,0.1)$.
In Sr$_2$RuO$_4$, $t'/t\sim0.1$ \cite{Sigrist,Nomura,Yanase}.
The electron density per spin, $n$, is set as $0.6$ and $1.4$.
(Note that $t,t'<0$ in the present Hubbard model according to 
Slater-Koster \cite{Slater}.
However, we assume $t,t'>0$ because they are positive in Sr$_2$RuO$_4$
due to the presence of $p_z$-orbital between the nearest $Ru$-sites 
\cite{Sigrist,Nomura,Yanase}.)
The splitting $\Delta^\pm$ represents the minimum band-splitting 
($|E_\k^+-E_\k^-|$) measured from the the Fermi surface of 
$E_\k^\pm$-band.
In Fig. \ref{fig:FS},
$\k^*$ represents the position of the minimum band-splitting,
 $\Delta \equiv \min \{\Delta^+,\ \Delta^- \}$.

Here, we consider the quasiparticle damping rate $\gamma=\hbar/2\tau$,
which is given by the imaginary part of the self-energy ${\hat \Sigma_\k(\e)}$.
For simplicity, we assume that $\gamma$ is diagonal with respect to orbital,
and independent of momentum.
Then, retarded and advanced Green functions with finite $\gamma$
are given by
\begin{eqnarray}
G_{\a\b'}^R(\w)&=& G_{\a\b'}(\w+i\gamma) \label{eqn:GR},
 \\
G_{\a\b'}^A(\w)&=& G_{\a\b'}(\w-i\gamma) \label{eqn:GA},
\end{eqnarray}
where $\a,\b=xz,yz$.
This assumption will be justified when the damping is caused by 
local impurities since the local Green function 
$g_{\a,\b}(\e)= n_{\rm imp}I^2 \sum_\k G_{\a,\b}(\k,\e)$ 
is diagonal and independent of $({\a,\b})$ if $\lambda=0$.
[Offdiagonal term of $g_{\a,\b}(\e)$ is proportional to $\lambda$.]
In the Born approximation,
${\hat \Sigma(\e)}= n_{\rm imp} I^2 {\hat g(\e)}$,
where $n_{\rm imp}$ is the impurity concentration
and $I$ is the impurity potential.
The Born approximation is valid when $I$ is much smaller than
the bandwidth $W_{\rm band}$. 
In Appendix \ref{AP-offd}, we derive the AHC in the Born approximation
by taking the offdiagonal term of ${\hat \Sigma(\e)}$ into account correctly,
and find that its effect is negligible.

On the other hand, 
the assumption given in eqs. (\ref{eqn:GR}) and (\ref{eqn:GA})
may not be valid in strongly correlated systems.
In Appendix \ref{AP-mom} and \ref{AP-band}, 
we study the intrinsic AHC when $\gamma$ depends on
momentum and band index, and find that the AHC could be changed
prominently by these influences.

\section{anomalous Hall conductivity}
 \label{AHE}
In this section, we derive the intrinsic AHC $\s_{xy}^a$ 
(hereafter, we drop the superscript $a$) and the 
longitudinal conductivity $\s_{xx}$ for finite quasiparticle
damping rate $\gamma$ based on the linear-response theory
 \cite{Nakano}.
Here, we take the self-energy correction due to correlation,
which causes finite quasiparticle damping rate.
On the other hand, we drop all the current vertex correction (CVC).
The CVC due to impurities with $ls$-coupling give rise to 
the skew scattering \cite{Bruno}.
However, CVC due to local impurities vanishes identically.
Note that the CVC due to the Coulomb interaction 
does not cause the skew scattering
 \cite{Kontani94}.
Hereafter, we put the renormalization factor $z=1$
because $z$ exactly cancels in the final formula of the AHC.

\begin{figure}[htbp]
\includegraphics[width=0.8\linewidth]{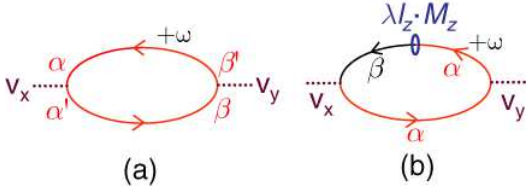}
\caption{(color online) (a) Diagrammatic expression for $\s_{xy}$.
(b) First order term with respect to the atomic $ls$-coupling
in terms of the band-diagonal representation.
}
  \label{fig:diagram}
\end{figure}

From now on, we assume a complete ferromagnetic state with
$\downarrow$-spin electrons only, i.e., ${\rm sgn}(s_z)=-1$.
Hereafter, we drop the factor $e^2/\hbar=2\pi e^2/h$
($h$ being the Plank constant) in $\s_{\mu\nu}$ to simplify expressions.
According to Streda \cite{Streda}, $\s_{\mu\nu}$ is given by
\begin{eqnarray}
\s_{\mu\nu}&=& \s_{\mu\nu}^I + \s_{\mu\nu}^{II} ,
 \\
\s_{\mu\nu}^I&=&
\sum_{\k,\a\a'\b\b'} \int\frac{d\e}{2\pi}
 v_\mu^{\a'\a} v_\nu^{\b'\b}
 \left( -\frac{\d f}{\d \e} \right) 
\nonumber \\
& &\times 
\left[ G_{\a\b'}^R G_{\b\a'}^A 
 -\frac12 \left( G_{\a\b'}^R G_{\b\a'}^R
+  G_{\a\b'}^A G_{\b\a'}^A \right) \right] ,
 \nonumber \\ 
 \label{eqn:sigma-I} \\
\s_{\mu\nu}^{II}&=&
-\frac{1}{2}\sum_{\k,\a\a'\b\b'} \int\frac{d\e}{2\pi}
 v_\mu^{\a'\a} v_\nu^{\b'\b} f(\w)
 \nonumber \\
& &\times 
 \left[ \frac{\d}{\d\e}G_{\a\b'}^R \cdot G_{\b\a'}^R
 -G_{\a\b'}^R \frac{\d}{\d\e} G_{\b\a'}^R \right. 
 \nonumber \\
& &\ \ \ \ \ \ 
 \left. -\frac{\d}{\d\e}G_{\a\b'}^A \cdot G_{\b\a'}^A
 +G_{\a\b'}^A \frac{\d}{\d\e} G_{\b\a'}^A \right] ,
 \label{eqn:sigma-II}
\end{eqnarray}
where $ v_\mu^{\a'\a}$ and $G_{\a\b'}$ are given by
eqs. (\ref{eqn:vx}), (\ref{eqn:vy}), (\ref{eqn:GR}) and (\ref{eqn:GA}).
$f(\e)$ is the Fermi distribution function.
The diagrammatic expression is given in fig. \ref{fig:diagram} (a).
We call $\s_{\mu\nu}^I$ and $\s_{\mu\nu}^{II}$
the ``Fermi surface term'' and the ``Fermi sea term'',
respectively, according to literature.

The original KL's paper \cite{KL} studied only a part of 
the Fermi sea term ($\s_{xy}^{IIb}$ given in eq. (\ref{eqn:xy-II2})),
which is equivalent to the Berry curvature term.
On the other hand, Kontani and Yamada \cite{Kontani94}
derived the Fermi surface term $\s_{xy}^{I}$ in the $J=5/2$
periodic Anderson model.
They also showed that the Fermi sea term is absent in this model.
In general models, however, both contributions exist as recognized 
by Luttinger \cite{KL2}.
Recently, a detailed analysis for both terms 
was given in ref. \cite{Mac} for graphene.
However, comparison of each term was not complete for $d$-electron systems.
In the present paper, 
we show that $\s_{xy}^{I}$ is much larger than $\s_{xy}^{II}$ 
in metallic systems for a wide range of $\gamma=\hbar/2\tau$.

We note that the normal Hall conductivity due to the 
Lorentz force is proportional to $\gamma^{-2}$.
In this case, $\s_{xy}^{I}$ is much larger than 
$\s_{xy}^{II} \sim O(\gamma^0)$ in good metals where 
$\tau= 1/2\gamma^{-1}$ is very large.
On the other hand, the intrinsic AHC is of the order of $\gamma^0$.
By this reason, both $\s_{xy}^{I}$ and $\s_{xy}^{II}$ 
could be the same order.

From now on, we calculate the intrinsic AHC for $(\mu,\nu)=(x,y)$.
First, we take the summation over $\a',\a,\b',\b$ in eqs.
(\ref{eqn:sigma-I}) and (\ref{eqn:sigma-II}).
After taking the $\k$-summation, only the terms
$(\a',\a,\b',\b)=(x,x,x,y), (x,x,y,x), (x,y,x,x), (y,x,x,x)$
remain finite. 
Using the square lattice symmetry of the present model, we obtain that 
\begin{eqnarray}
\s_{xy}^I&=&
\l \sum_{\k} \int\frac{d\e}{2\pi}  v_x^{xx} v_y^{xy} 
 \left( -\frac{\d f}{\d \e} \right) \frac{4\g}{d^Rd^A} ,
 \label{eqn:sxy-I} \\
\s_{xy}^{II}&=&
2i\l \sum_{\k} \int\frac{d\e}{2\pi}
  v_x^{xx} v_y^{xy} f(\w)
 \nonumber \\
& &\times \left[ \left(\frac1{d^R}\right)^2
- \left(\frac1{d^A}\right)^2 \right] ,
 \label{eqn:sxy-II}
\end{eqnarray}
where $v_x^{xx} v_y^{xy}= 8tt'\sin^2k_x \cos k_y$.
Here, it is essential that $v_x^{xx} v_y^{xy}$ is the
totally symmetric representation ($A_{1g}$),
and it does not vanish after the $\k$-summation.
Note that the term proportional to $G^RG^R+G^AG^A$
in $\s_{xy}^I$ vanishes identically.

Here, we derive the expressions for $\s_{xy}^I$ and $\s_{xy}^{II}$ at $T=0$.
Then, the $\w$-integration in eq. (\ref{eqn:sxy-II})
can be performed using the relation
\begin{eqnarray}
\int_{-\infty}^\mu \frac{dx}{(x-a)^2(x-b)^2}
&=& \frac{-(2\mu-a-b)}{(a-b)^2(\mu-a)(\mu-b)}
\nonumber \\
& & - \frac2{(a-b)^3} \ln \left(\frac{a-\mu}{b-\mu}\right) .
\end{eqnarray}
The final result for the intrinsic AHC is given by
\begin{eqnarray}
\s_{xy}&=& \s_{xy}^{I} + \s_{xy}^{IIa} + \s_{xy}^{IIb},
 \label{eqn:xy} \\
\s_{xy}^{I}&=&
\frac{2\l}{\pi} \sum_\k v_x^{xx}v_y^{xy}
 \nonumber \\
& &\times \frac{\gamma}{((\mu-E_\k^+)^2+\gamma^2)((\mu-E_\k^-)^2+\gamma^2)} ,
 \label{eqn:xy-I} \\
\s_{xy}^{IIa}&=&
\frac{2\l}{\pi} \sum_\k v_x^{xx}v_y^{xy}
 \frac1{(E_\k^+ -E_\k^-)^2}
 \nonumber \\
& &\times {\rm Im} \left\{ \frac{2\mu-E_\k^+ -E_\k^- +2i\gamma}
 {(\mu-E_\k^+ +i\gamma)(\mu-E_\k^- +i\gamma)} \right\} ,
 \label{eqn:xy-II1} \\
\s_{xy}^{IIb}&=&
\frac{4\l}{\pi} \sum_\k v_x^{xx}v_y^{xy}
 \frac1{(E_\k^+ -E_\k^-)^3} 
 \nonumber \\
& &\times {\rm Im} \left\{ 
 \ln \left( \frac{E_\k^+ -\mu -i\gamma}{E_\k^- -\mu -i\gamma}
 \right) \right\} .
 \label{eqn:xy-II2}
\end{eqnarray}
Below, we perform the numerical calculation for the AHC
using eqs. (\ref{eqn:xy-I}), (\ref{eqn:xy-II1}) 
and (\ref{eqn:xy-II2}).

In the same way, we calculate the longitudinal conductivity
for $\mu=\nu=x$.
In this case, (\ref{eqn:sigma-II}) vanishes identically,
and only the Fermi surface term (\ref{eqn:sigma-I}) remains finite. 
After taking the $\k$-summation and using the square lattice symmetry
of the present model, we obtain the expression for $\s_{xx}$ at $T=0$ 
as follows:
\begin{eqnarray}
\s_{xx}&=& \s_{xx}^{Ia} + \s_{xx}^{Ib} ,
 \label{eqn:xx} \\
\s_{xx}^{Ia}&=&
\frac{1}{2\pi} \sum_\k \left\{ (v_x^{xx})^2[ (\mu-\xi_y)^2+\gamma^2 ]
 \right. + 2(v_x^{xy})^2 \nonumber \\
& &\ \ \times [(\xi^{xy})^2 - \l^2 
   + (\mu-\xi_y)(\mu-\xi_x)+\gamma^2]
 \nonumber \\
& &\left. + 4v_x^{xx}v_x^{xy} \xi^{xy}(\mu-\xi_y) \right\}
 /d^R(0)d^A(0) ,
 \label{eqn:xx-I1} \\
\s_{xx}^{Ib}&=&
\frac{-1}{2\pi} {\rm Re} \sum_\k 
 \left\{ (v_x^{xx})^2(\mu-\xi_y +i\gamma)^2 \right.  
 + 2(v_x^{xy})^2 \nonumber \\
& &\ \times \left[ (\xi^{xy})^2 - \l^2 
 + (\mu-\xi_y+i\gamma)(\mu-\xi_x+i\gamma) \right] 
 \nonumber \\
& &\left. + 4v_x^{xx}v_x^{xy} \xi^{xy}(\mu-\xi_y+i\gamma) \right\}
 /(d^R(0))^2  .
 \label{eqn:xx-I2} 
\end{eqnarray}
Note that $\s_{xx}^{II}$ vanishes identically. 
In the case of $\gamma\rightarrow 0$,
$\s_{xx}^{Ia}$ diverges as $\gamma^{-1}$ whereas $\s_{xx}^{Ib}$ is finite. 
Therefore, $\s_{xx} \sim \s_{xx}^{Ia}$ in good metals.
However, $\s_{xx}^{Ia}$ and $\s_{xx}^{Ib}$ become the same order
when $\gamma$ is very large. 
We perform the numerical calculation for the $\s_{xx}$
using eqs. (\ref{eqn:xx-I1}) (\ref{eqn:xx-I2}) in later sections.

Before performing the numerical study in \S \ref{Numerical},
we analyze eqs. (\ref{eqn:xy-I}), (\ref{eqn:xy-II1}) 
and (\ref{eqn:xy-II2}) when $\gamma$ is very small in detail:
In this case,
\begin{eqnarray}
& &\frac1{d^R(0)d^A(0)}
 \approx \frac{\pi}{\gamma}
 \frac{\delta(\mu-E_\k^+)+\delta(\mu-E_\k^-)}
 {(E_\k^+ -E_\k^-)^2+\gamma^2} ,
 \\
& &{\rm Im} \left\{ \frac{2\mu-E_\k^+ -E_\k^- +2i\gamma}
 {(\mu-E_\k^+ +i\gamma)(\mu-E_\k^- +i\gamma)} \right\}
 \nonumber \\
& &\ \ \ \ \ \ \ \ \ \ \ \ \ \ 
 \approx -\pi\delta(\mu-E_\k^+) -\pi\delta(\mu-E_\k^-) ,
 \\
& &{\rm Im} \left\{ \ln \left( 
 \frac{E_\k^+ -\mu -i\gamma}{E_\k^- -\mu -i\gamma}
 \right) \right\}
 \nonumber \\
& &\ \ \ \ \ \ \ \ \ \ \ \ \ \ 
 \approx -\pi\theta(\mu-E_\k^+) +\pi\theta(\mu-E_\k^-) .
\end{eqnarray}
Substituting above equations into (\ref{eqn:xy-I})-(\ref{eqn:xy-II2}),
we obtain the following relation for $\gamma\rightarrow 0$:
\begin{eqnarray}
\s_{xy}^{I}
 &\approx& 2\l\sum_\k v_x^{xx} v_y^{xy}
 \nonumber \\
& &\times \frac{\delta(\mu-E_\k^+)+\delta(\mu-E_\k^-)}
 {(E_\k^+ - E_\k^-)^2} ,
 \label{eqn:ap-xy-I} \\
\s_{xy}^{IIa}
 &\approx& -\s_{xy}^{I} ,
 \label{eqn:ap-xy-II1} \\
\s_{xy}^{IIb}
 &\approx& 4\l\sum_\k v_x^{xx} v_y^{xy}
 \nonumber \\
& &\times \frac{-\theta(\mu-E_\k^+)+\theta(\mu-E_\k^-)}
 {(E_\k^+ - E_\k^-)^3} .
 \label{eqn:ap-xy-II2}
\end{eqnarray}
According to eq. (\ref{eqn:ap-xy-II2}),
the main contribution to $\s_{xy}^{IIb}$ comes from an area
near $\k^*$ in fig \ref{fig:FS}, if $E_\k^+>0$ and $E_\k^-<0$ are satisfied.
When $d_\k \equiv E_\k^+-E_\k^-$ is small, then
$(-\theta(\mu-E_\k^+)+\theta(\mu-E_\k^-))/d_\k \approx 
2\delta(\mu-(E_\k^++E_\k^-)/2)$.
In this case,
\begin{eqnarray}
\s_{xy}^{IIb}
 &\approx& 4\l\sum_\k v_x^{xx} v_y^{xy}
 \frac{\delta(\mu-(E_\k^++E_\k^-)/2)}
 {(E_\k^+ - E_\k^-)^2}
 \nonumber \\
&\approx& \s_{xy}^{I} .
 \label{eqn:approx}
\end{eqnarray}
Note that this relation will not be satisfied in a special case
where the Fermi level lies between a narrow bandgap.
In the next section, we see that eq. (\ref{eqn:approx}) holds very well
in the present model.

Here, we summarize the obtained results in this section.\\
(i) $\s_{xy}^{IIb}$ was first recognized by Karplus and Luttinger
 \cite{KL}.
It can be rewritten in terms of the summation of the Berry curvature
 \cite{MOnoda}.\\
(ii)$\s_{xy}^{IIa}$ is another ``Fermi sea term'', although
its final expression (\ref{eqn:ap-xy-II1}) is written like
a Fermi surface term.
We find that $\s_{xy}^{II}$ is very small in usual metals because
$\s_{xy}^{IIa}$ and $\s_{xy}^{IIb}$ almost cancels each other.\\
(iii) The Fermi surface term is canceled by $\s_{xy}^{IIa}$
in the clean limit, which is also recognized
in 2D Dirac model \cite{Sinitsyn2}.
However, this cancellation is imperfect when $\gamma$ is finite,
as shown in the next section. \\
Several previous works assumed that the intrinsic AHC is 
given by $\s_{xy}^{IIb}$
 \cite{MOnoda,Fe,Fang}.
One might consider that this assumption is justified 
by the present calculation for $\gamma\rightarrow 0$.
However, this assumption is not always guaranteed 
as we will see in the next section:
When $\gamma$ is as large as $\Delta$, then 
(a) $\s_{xy}\approx \s_{xy}^{I}$ still holds whereas 
(b) $\s_{xy}$ becomes quite different from $\s_{xy}^{IIb}$.
In Appendix \ref{AP-mom} and \ref{AP-band}, 
we study the intrinsic AHC in the general case where
$\gamma$ is $\k$-dependent or it depends on band index.
In these cases, statement (a) and (b) are also true.
Therefore, overall behavior of $\s_{xy}$ for a wide range of $\gamma$
is well expressed by the Fermi surface term $\s_{xy}^{I}$, 
not by $\s_{xy}^{IIb}$.
This is an important result of this paper, which is obtained 
by considering all the terms contributing to the intrinsic AHC.

Finally, we comment that both $t'$ and ${\hat h}^\l$
change their signs under the gauge transformation
$|xz\rangle \rightarrow -|xz\rangle$.
The AHC is invariant under this gauge transformation.

\section{Numerical Study}
 \label{Numerical}
In this section, we perform the numerical calculation for both
$\s_{xy}$ and $\s_{xx}$ at $T=0$, assuming a complete 
ferromagnetic state where $n_\downarrow=n$ and $n_\uparrow=0$.
In this case, $m_z= \mu_{\rm B}n$. (We put $\mu_{\rm B}=1$ hereafter.)
The main purpose of this section is to 
elucidate both the filling ($n$) and damping rate
($\gamma$) dependences of the AHC.
We perform the $\k$-summations in eq. (\ref{eqn:xy})
for $\s_{xy}$ and eq. (\ref{eqn:xx}) for $\s_{xx}$
numerically, dividing the Brillouin zone into 5000$\times$5000 meshes.

The unit of conductivity in this section is $e^2/ha$,
where $h$ is the Plank constant and $a$ is the unit cell length
(inter-layer distance in 2D systems).
If we assume the length of unit cell $a$ is 4\AA,
$e^2/ha \approx 10^3 \Omega^{-1}{\rm cm}^{-1}$.
\begin{figure}[htbp]
\includegraphics[width=.9\linewidth]{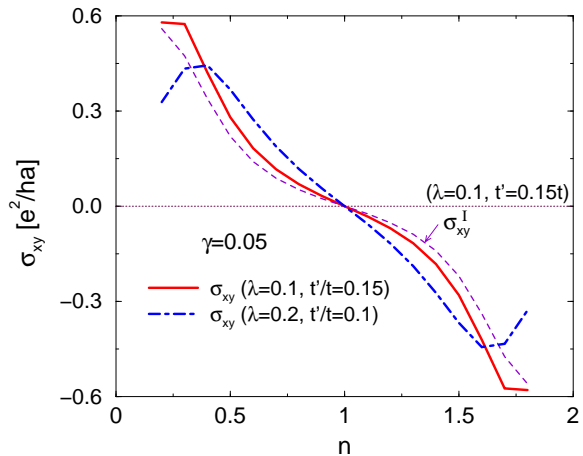}
\caption{(color online) Obtained total AHC ($\s_{xy}$) and $\s_{xy}^I$
as a function of $n=m_z$.
It is shown that $\s_{xy} \approx \s_{xy}^I$ is well satisfied.
$e^2/ha \approx 10^3 \Omega^{-1}{\rm cm}^{-1}$ if we put 
the unit cell length $a=4$\AA.
$\s_{xy}$ is not a monotonic function of $m_z$,
and changes its sign at $m_z \sim 1$
}
  \label{fig:AHE-n}
\end{figure}

\begin{figure}[htbp]
\includegraphics[width=.8\linewidth]{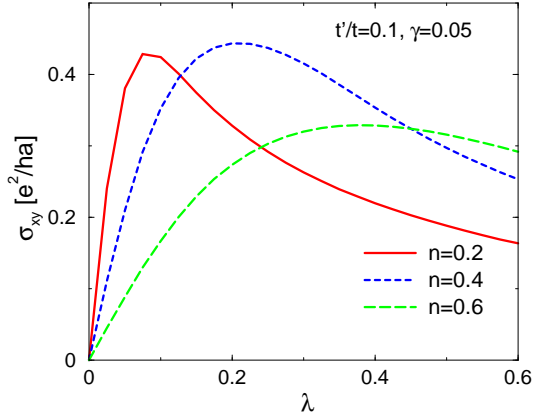}
\caption{(color online) $\lambda$-dependence of $\s_{xy}$ for 
$n=1.4$ and $0.6$.
}
  \label{fig:lambda}
\end{figure}

Figure \ref{fig:AHE-n} shows the $n$-dependence of $\s_{xy}$
for $n=0.2\sim1.8$. $n=2$ corresponds to a ferromagnetic band insulators. 
In the present model, $\s_{xy}(n)=-\s_{xy}(2-n)$.
Here, we put $\gamma=0.05$, which is sufficiently smaller than
the minimum band-splitting measured from the Fermi surface, $\Delta$,
which is shown in Fig. \ref{fig:FS}.
The bandwidth of the present model is approximately $4|t|=4$.
We put the $ls$-coupling constant $\lambda$ as $0.1\sim0.2$.
If we assume $t=4000$K, it corresponds to $\lambda=400\sim800$K,
which are realistic values in transition metals.
In the present study, $|\s_{xy}|$ exceeds 
$0.5\times10^3 \Omega^{-1}{\rm cm}^{-1}$ at $n\sim 1.8$.
The obtained magnitude of $|\s_{xy}|$ is comparable
with experimental value in Fe \cite{Fe} and in SrRuO$_3$ \cite{Fang}.
Figure \ref{fig:lambda} shows the $\l$-dependence of $|\s_{xy}|$
for $n=0.6$ and 1.4.
$|\s_{xy}|$ is approximately proportional to $\l$ below $\l\sim0.2$,
whereas it tends to saturate when $\l$ is as large as $\Delta_{\l=0}$
since $\Delta= E_{\k^*}^+ -E_{\k^*}^-$ increases with $\l$.
 \cite{Fe}.

In Fig. \ref{fig:AHE-n},
$\s_{xy}$ is positive (negative) for $n<1$ ($n>1$);
$\s_{xy}=0$ at $n\approx 1$.
The reason for the sign change of $\s_{xy}=0$ is the following:
According to eq. (\ref{eqn:xy}), the main contribution
for $\s_{xy} \ (\approx \s_{xy}^I \ {\rm or} \  \s_{xy}^{IIb})$
comes from an area near $k^*$. 
Therefore, $\s_{xy} \sim \left. v_x^{xx}v_y^{xy} \right|_{\rm k^*}
 \sim tt' \sin^2 k_x^* \cos k_y^*$.
Because $tt'>0$, the sign of $\s_{xy}$ changes to be negative
when $|k_x^*|=|k_y^*|$ exceeds $\pi/2$.
The sign change of the AHC as a function of $m_z$ is
actually observed in SrRuO$_3$.

\begin{figure}[htbp]
\includegraphics[width=.9\linewidth]{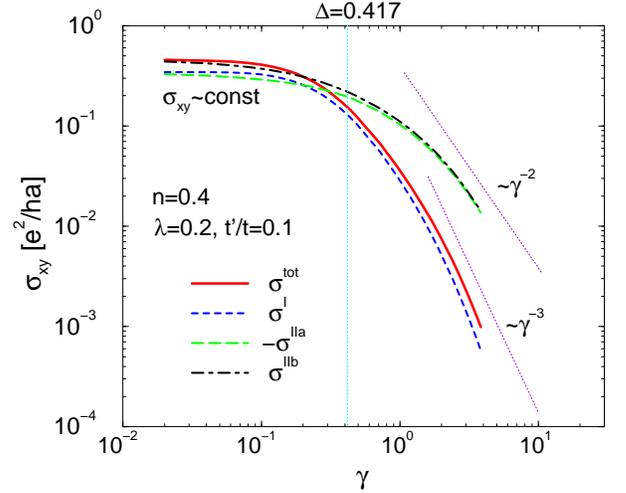}
\caption{(color online) $\g$-dependences of $\s_{xy} \ (=\s^{\rm tot})$, 
$\s_{xy}^I \ (=\s^{I})$, $\s_{xy}^{IIa} \ (=\s^{IIa})$ 
and $\s_{xy}^{IIb} \ (=\s^{IIb})$, respectively.
$\s_{xy}$ shows a crossover behavior at $\g\sim\Delta$.
It is shown that $\s_{xy}^I$ reproduces a overall behavior
of the total $\s_{xy}$ for a wide range of $\gamma$.
}
  \label{fig:logsxy-123}
\end{figure}

Next, we examine the $\gamma$-dependence of the AHC.
Figure \ref{fig:logsxy-123} shows the total AHC ($\s_{xy}$),
$\s_{xy}^I$, $-\s_{xy}^{IIa}$ and $\s_{xy}^{IIb}$ for $n=1.4$.
We see that all of them are $\gamma$-independent when $\gamma\ll \Delta$.
On the other hand, they start to decrease with $\gamma$ 
when $\gamma \gg \Delta$:
This is easily recognized from the functional forms of eqs.
(\ref{eqn:xy-I})-(\ref{eqn:xy-II2}).
We see that $\s_{xy}^I$ is almost equal to $-\s_{xy}^{IIa}$
and $\s_{xy} \approx \s_{xy}^{IIb}$ when $\gamma\ll \Delta$, 
as discussed in the previous section.
On the other hand, $\s_{xy} \approx \s_{xy}^I$ whereas
$\s_{xy}$ is quite different from $\s_{xy}^{IIb}$ for $\gamma\gg \Delta$.
As a result, the Fermi surface term $\s_{xy}^{I}$
succeeds in reproducing the overall behavior of the AHC 
for a wide range of $\gamma$.
$\s_{xy}$ is proportional to $\gamma^{-3}$ for $\gamma\sim W_{\rm band}$,
which can be recognized by eq. (\ref{eqn:xy-I}).

This crossover behavior of $\s_{xy}$ at $\gamma\sim \Delta$
was first pointed out by ref. \cite{Kontani94}.
They found that $\s_{xy}$ is proportional to $\gamma^{-2}$
for $|E^f-\mu| \ll \gamma$ in $J=5/2$ periodic Anderson model (PAM).
($E^f$ is the energy of $f$-electrons.)
The different $\gamma$-dependence of $\s_{xy}$ between in $d$-$p$ model
and in PAM comes from the fact that the conduction bandwidth 
$W_{\rm band}^c$ is much wider than the heavy quasiparticle band in PAM.
In the PAM, the quasiparticle damping is given by
$\g= \Gamma \cdot V^2/((\mu-E^f)^2+\Gamma^2)$, where $V$ is the 
$c$-$f$ mixing term and $\Gamma$ is the imaginary part of the 
$f$-electron self-energy.
Because $\g$ saturates at $\Gamma\sim |\mu-E^f|$ and
the relation $V^2/(\mu-E^f)\sim W_{\rm band}^c$ holds in the PAM,
$\g$ will always be smaller than $W_{\rm band}^c$
even if $\Gamma \gg W_{\rm band}^c$.

\begin{figure}[htbp]
\includegraphics[width=.9\linewidth]{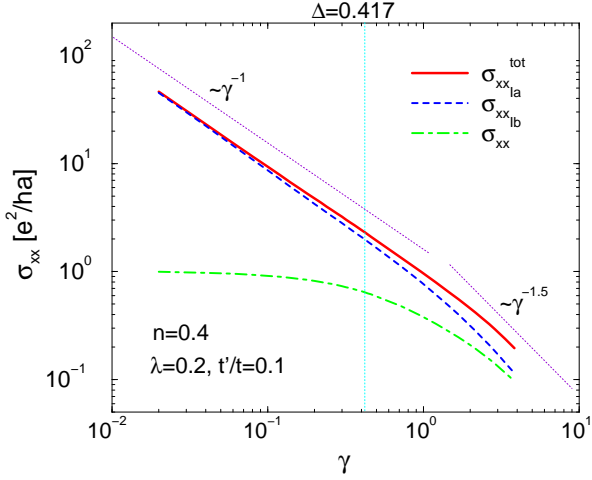}
\caption{(color online) $\g$-dependences of $\s_{xx}^{Ia}$, $\s_{xx}^{Ib}$
and $\s_{xx}= \s_{xx}^{Ia} + \s_{xx}^{Ib}$.
}
  \label{fig:logsxx-12}
\end{figure}

We also study the $\gamma$-dependences of the longitudinal 
conductivity $\s_{xx}$.
Figure \ref{fig:logsxx-12} shows that the relations
$\s_{xx} \propto \gamma^{-1}$ and $\s_{xx}^{Ia}\gg\s_{xx}^{Ib}$
hold well for a wide range of $\gamma$.
The conductivity $\s_{xx}$ does not show any clear crossover 
behavior at $\gamma\sim \Delta$.
When $\gamma\sim W_{\rm band}$, $\s_{xx}$ seems 
proportional to $\gamma^{-1.5}$, and $\s_{xx}^{Ib}$ is 
as large as $\s_{xx}^{Ia}$.
Although eqs. (\ref{eqn:xx-I1}) and (\ref{eqn:xx-I2})
suggest that $\s_{xx} \propto \gamma^{-2}$ when $\gamma\gg W_{\rm band}$,
we could not find such a behavior for $\gamma \gtrsim W_{\rm band}$.

\begin{figure}[htbp]
\includegraphics[width=.9\linewidth]{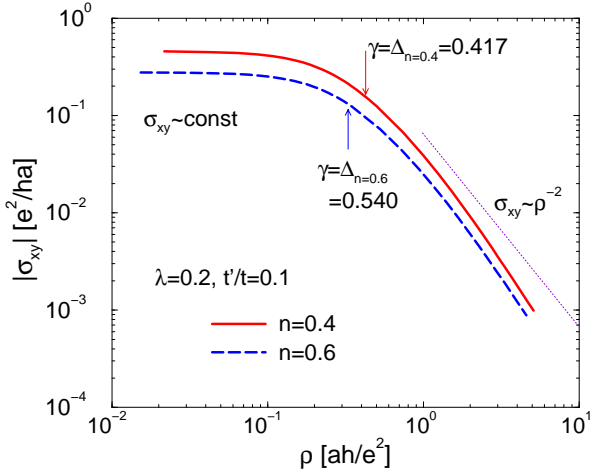}
\caption{(color online) 
Obtained relation between $\s_{xy}$ and $\rho=1/\s_{xx}$
for $n=1.6$ and $0.4$.
$\rho=0.1$ corresponds to $100\mu\Omega{\rm cm}$. 
It reproduces well the experimental universal 
``Asamitsu plot'' for various transition metal ferromagnets, 
where $\s_{xy} =10^2\sim10^3 \Omega^{-1}{\rm cm}^{-1}$
for $\rho=1\sim 100\mu\Omega{\rm cm}$, whereas
$\s_{xy} \propto \rho^{-2}$ for $\rho \gg 100\mu\Omega{\rm cm}$.
}
  \label{fig:logsxy}
\end{figure}

Finally, we study the relation between the AHC and 
the resistivity $\rho=1/\s_{xx}$.
Figure \ref{fig:logsxy} shows that $\s_{xy}$ is independent 
of $\rho$ when $\rho\lesssim 0.1 \ [\sim 100\mu\Omega{\rm cm}]$.
On the other hand, $\s_{xy}$ starts to decrease 
for $\rho\gtrsim 0.1$ in proportion to $\rho^{-2}$.
This crossover behavior of $\s_{xy}$ at 
$\rho\sim 100\mu\Omega{\rm cm}$ is observed universally by
recent experiments on various transition metal ferromagnets
by Asamitsu et al.\cite{Asamitsu}.

\section{Analysis of anomalous Hall conductivity 
in the Band-Diagonal Representation}
 \label{BandDiagonal}
In this section, we analyze the intrinsic AHC in terms of the band-diagonal 
representation, which has been analyzed by many authors.
We note that this representation is useful for the analysis of the AHC
only when $\gamma_\a$ is orbital independent.
In this representation, 
according to eqs. (\ref{eqn:sigma-I}) and (\ref{eqn:sigma-II}),
\begin{eqnarray}
\s_{xy}^I&=&
 \sum_{\k,l\ne m} \frac{1}{2\pi}  v_x^{m l} v_y^{l m} 
 \frac1{(\mu-E_\k^l+i\gamma)(\mu-E_\k^m-i\gamma)} ,
 \nonumber \\
 \label{eqn:AP-I} \\
\s_{xy}^{II}&=&
 i\sum_{\k,l\ne m} \int_{-\infty}^\mu \frac{d\e}{2\pi}
  v_x^{m l} v_y^{l m}
 \nonumber \\
& &\times {\rm Im}\left\{
  \frac1{(\e-E_\k^l+i\gamma)^2} \frac1{(\e-E_\k^m+i\gamma)} \right.
 \nonumber \\
& &\ \ \ \ \ \ \left.
 -\frac1{(\e-E_\k^l+i\gamma)} \frac1{(\e-E_\k^m+i\gamma)^2} \right\} ,
 \label{eqn:AP-II}
\end{eqnarray}
at zero temperature.
Here, $l$ and $m$ are the band suffices, and we dropped
the diagonal terms $l=m$ because they vanish identically.
After performing the $\e$-integral in $\s_{xy}^{II}$,
the Fermi sea term is expressed as $\s_{xy}^{IIa}+\s_{xy}^{IIb}$:
\begin{eqnarray}
\s_{xy}^{IIa}&=&
\frac{i}{2\pi} \sum_{\k,l\ne m} v_x^{m l}v_y^{l m}
 \frac1{E_\k^l -E_\k^m}
 \nonumber \\
& &\times {\rm Im} \left\{ \frac{E_\k^l +E_\k^m -2\mu -2i\gamma}
 {(E_\k^l -\mu+ -i\gamma)(E_\k^m -\mu -i\gamma)} \right\} ,
 \label{eqn:AP-II1} \\
\s_{xy}^{IIb}&=&
\frac{-i}{\pi} \sum_{\k,l\ne m} v_x^{m l}v_y^{l m}
 \frac1{(E_\k^l -E_\k^m)^2} 
 \nonumber \\
& &\times {\rm Im} \left\{ 
 \ln \left( \frac{E_\k^l -\mu -i\gamma}{E_\k^m -\mu -i\gamma}
 \right) \right\} .
 \label{eqn:AP-II2}
\end{eqnarray}
When $\gamma$ is very small, they are given by
\begin{eqnarray}
\s_{xy}^I&=&
 \frac{-i}{2}\sum_{\k,l\ne m} \delta(\mu-E_\k^l)
 \frac{ v_x^{m l} v_y^{l m} - v_x^{l m} v_y^{m l} }{E_\k^l-E_\k^m},
 \nonumber \\
 \label{eqn:APP-I} \\
\s_{xy}^{IIa}&=& -\s_{xy}^I ,
 \label{eqn:APP-II1} \\
\s_{xy}^{IIb}&=& i\sum_{\k,l\ne m}f(E_\k^l) 
 \frac{v_x^{m l} v_y^{l m}-v_x^{l m} v_y^{m l}}
 {(E_\k^l-E_\k^m)^2} .
 \label{eqn:APP-II2} 
\end{eqnarray}
%
Because $v_\mu^{l m}= J_\mu^{lm}/(E_\k^m-E_\k^l)$ and 
$J_\mu^{lm} = \langle \k,l|\d/\d k_\mu |\k,m\rangle$,
$\s_{xy}^{IIb}$ is equivalent to eq. (2.16) 
of ref. \cite{KL}, or eq. (3) of ref. \cite{MOnoda}.
Based on eq. (\ref{eqn:APP-II2}),
When the Fermi energy lies in a gap, $\s_{xy}^{IIb}$ is 
quantized when the Fermi energy lies in a gap \cite{Thouless}.
In the metallic state, on the other hand,
$\s_{xy}^{IIb}$ is rewritten as the Fermi surface term
due to the partial integral.

Off course, we can reproduce eqs. 
(\ref{eqn:ap-xy-I})-(\ref{eqn:ap-xy-II2}) by applying
eqs. (\ref{eqn:APP-I})-(\ref{eqn:APP-II2}) to
the present tight-binding model given in \S \ref{Model}.
In the band-diagonal representation ($E_\k^\a; \ \a=\pm$), 
we can show that the off-diagonal velocity is given by
\begin{eqnarray}
v_x^{+-}&=&-\left\{ |\a|^2v_x^{xx}-\xi^{xy}(\xi^x-\xi^y)v_x^{xy} \right.
 \nonumber \\
& & \left. i\l v_x^{xy}|E_\k^+-E_\k^-| \right\} 
 /(|\a||E_\k^+-E_\k^-|) ,
 \\
v_y^{+-}&=&\left\{ |\a|^2v_y^{yy}-\xi^{xy}(\xi^x-\xi^y)v_y^{xy} \right.
 \nonumber \\
& & \left. -i\l v_y^{xy}|E_\k^+-E_\k^-| \right\} 
 /(|\a||E_\k^+-E_\k^-|) ,
\end{eqnarray}
and $v_\mu^{-+}= \{v_\mu^{+-}\}^*$.
Therefore,
\begin{eqnarray}
v_x^{\a{\bar \a}}v_y^{{\bar \a}\a}&=&
-i\l \frac{ v_x^{xx}v_y^{xy}+v_x^{xy}v_y^{yy}}{E_\k^\a-E_\k^{\bar \a}}
 \nonumber \\
& &+ ({\rm non \ A_{1g} \ terms}) ,
\end{eqnarray}
where $\a=\pm$ and ${\bar \a}=-\a$.
Substituting this result into eqs. (\ref{eqn:APP-I})-(\ref{eqn:APP-II2}),
we obtain the same results given in \S \ref{AHE}.

\section{Discussions}
 \label{summary}
\subsection{Summary of the Present Study}

In the present paper,
we studied the mechanism of the intrinsic AHE in transition metal
ferromagnets based on the $(d_{xz},d_{yz})$-orbital tight-binding model.
The origin of the anomalous velocity is the 
hopping integral between $d_{xz}$ and $d_{yz}$ orbitals, $t'$.
By virtue of it, we could reproduce a typical experimental 
value of the AHC in ferromagnets; 
$10^2 \sim 10^3 \Omega^{-1}{\rm cm}^{-1}$.
Thus, the anomalous velocity due to atomic $d$-orbitals
will be the main origin of the AHE in transition metal oxides.
This fact has been overlooked in previous theoretical works
based on electron gas models without atomic orbitals.
We note that the present model breaks the parity symmetry 
[$(x,y)\rightarrow (-x,y)$] due to $t'\ne0$.
This is a necessary condition for a spontaneous Hall effect
 \cite{Haldane,MOnoda}.

In accord with the present study, 
paramagnetic compound Ca$_{1.7}$Sr$_{0.3}$RuO$_4$ 
shows large AHE under the magnetic field \cite{AHE-CSRO}.
Its magnitude is comparable with the large AHE in $f$-electron systems 
(such as UPt$_3$) which originates from the anomalous velocity due to 
atomic $f$-orbitals\cite{Kontani94}.
The Fermi surfaces of this compound are composed of 
$t_{2g}$-orbital ($d_{xz}$, $d_{yz}$, $d_{xy}$-orbitals);
$\gamma$-band composed of $d_{xy}$-orbital
is absent in the present ($d_{xz}$, $d_{yz}$)-tight-binding model.
We will study the AHE of this compound in more detail
based on the $t_{2g}$-orbital tight-binding model,
which reproduces the bandstructure of (Ca,Sr)$_2$RuO$_4$ accurately
 \cite{future}.

We derive a general expression for the AHC valid for 
any quasiparticle damping rate $\g$, by performing the 
analytic continuation carefully.
[Equations (\ref{eqn:xy-I}), (\ref{eqn:xy-II1}) 
and (\ref{eqn:xy-II2}) for $(d_{xz},d_{yz})$-orbital model,
and eqs. (\ref{eqn:AP-I}), (\ref{eqn:AP-II1}) and (\ref{eqn:AP-II2})
for a general model.]
Using the general expression,
we succeeded in explaining the experimental crossover
behavior of the AHC in bad metals; $\s_{xy}$ is constant for 
$\rho\lesssim 100\mu\Omega{\rm cm}$, whereas $\s_{xy}\propto \rho^{-2}$
for higher resistivity.
This overall behavior is mainly given by the 
Fermi surface term, $\s_{xy}^I$,
whose importance was intensively studied in ref. \cite{Kontani94}.

We stress that the intrinsic AHC in the present model is not a
monotonic function with respect to $m_z= n_\downarrow$.
In partial ferromagnets, the total AHC is given by
\begin{eqnarray}
\s_{xy}(n_\uparrow,n_\downarrow)= 
 \s_{xy}(\mu_\downarrow)- \s_{xy}(\mu_\uparrow)
\end{eqnarray}
where $\s_{xy}(\mu)$ is given by eq. (\ref{eqn:xy}),
and $\mu_{\uparrow(\downarrow)}$ is the chemical potential 
for the electron density $n_{\uparrow(\downarrow)}$.
We found that the sign-change of $\s_{xy}$ occurs in correspondence 
with sign of $v_x^{xy}$ at $\k^*$.
We expect that this result will explain the sign-change of the AHE
in SrRuO$_3$ as a function of the magnetization.
We comment that authors of ref. \cite{MOnoda} found that 
a large value of AHC appears from $\s_{xy}^{IIb}$
when Fermi level lies inside a narrow "anticrossing band gap'',
and they suggested that this mechanism accounts for 
a huge AHC in ferromagnets.
However, this condition will be satisfied only for a narrow range
of magnetization.
The present model with $(d_{xz},d_{yz})$-orbitals
can give an enough magnitude of the AHC, although band crossings 
are absent (except at $\k=(0,0)$ and $(\pi,\pi)$).
We stress that the AHC in the present model shows a sign change 
with respect to $m_z$, even if the topology of the Fermi surface 
is unchanged; see Fig. \ref{fig:AHE-n}.

The relation $\s_{xy}=\s_{xy}^{IIb}$ (Berry curvature term)
will not hold in usual metallic compounds:
For example, in bad metals where $\gamma$ is as large as the 
minimum band-splitting measured from the Fermi surface, $\Delta$,
the relations $\s_{xy} \approx \s_{xy}^{I}$ (Fermi surface term) 
and $|\s_{xy}^{I}| \gg |\s_{xy}^{II}|$ hold well for 
a wide range of $\gamma$ as shown in fig. \ref{fig:logsxy-123}.
Even in good metals, the relation $\s_{xy}=\s_{xy}^{IIb}$ 
is also invalid when (a) $\gamma$ is $\k$-dependent, and/or 
(b) $\gamma$ depends on band index, i.e., $\gamma_l/\gamma_m\ne1$:
In these cases, the Fermi surface term deviates from 
eq. (\ref{eqn:ap-xy-I}) as shown in Appendix \ref{AP-mom} and \ref{AP-band}.

In summary,
the relation $\s_{xy}=\s_{xy}^{IIb}$ will hold only in good metals
as well as when $\gamma_l(\k,\w)$ is independent of $l$, $\k$ and $\w$
On the other hand, the relation
\begin{eqnarray}
\s_{xy}\approx \s_{xy}^{I} \ \ \ {\rm(Fermi \ surface \ term)}
\end{eqnarray}
will be universal in real metallic systems,
since the Fermi sea terms almost cancel each other
except for a special situation where $\mu$ lies 
inside a narrow anticrossing band gap.
This result is consistent with recent theoretical work on 
metallic graphene \cite{Mac}.
For a quantitative study of the intrinsic AHC, however,
we have to calculate $\s_{xy}^{I}$, $\s_{xy}^{IIa}$ and 
$\s_{xy}^{IIb}$ on the same footing.
This is also important conclusion in this paper.

Finally, we comment that the present $(d_{xz},d_{yz})$-tight binding model
in a paramagnetic state shows a finite spin Hall conductivity (SHC)
$\s_{xy}^z$, that is, $\s_z$-spin current along $y$-axis occurs
under the electric field along $x$-axis.
$\s_{xy}^z$ is given by $(-\hbar/e)$ times eqs. (\ref{eqn:xy}).
As shown in Fig. \ref{fig:AHE-n},
$\s_{xy}^z$ reaches $0.6 \ [(\hbar/e)(e^2/ha)]
\sim 600 \ [\hbar e^{-1} \cdot \Omega^{-1}{\rm cm}^{-1}]$ 
for $a=$4\AA, which is almost one order of magnitude 
larger than typical SHC in semimetals
 \cite{Murakami}.
In later publications, we will study the spin Hall effect
in more detail
 \cite{future}

\subsection{Comparison with Experiments: Transition Metals}
Asamitsu et al. examined the experimental relation between 
the AHC and the resistivity $\rho$ in various ferromagnets.
They observed the dissipation-less intrinsic AHC
in the intermediate conducting region with 
$\rho=1\sim 100\mu\Omega{\rm cm}$.
In the bad metal region, $\s_{xy}$ is proportional to $\rho^{-2}$,
which is well explained in the present study.
We found this crossover at $\sim 100\mu\Omega{\rm cm}$
occurs when $\gamma\sim\Delta$.
According to Fig. \ref{fig:logsxx-12}, $100\mu\Omega{\rm cm}$ 
corresponds to $\Delta \sim 0.2 \approx 800$K.
Another crossover to the good metal region 
($\sim 1\mu\Omega{\rm cm}$) seems to be rather complex:
It might be due to the extrinsic AHE or a phenomenon
related to the normal Hall effect.

Recently, S. Onoda et al. \cite{SOnoda} calculated both the 
intrinsic AHC \cite{KL,Kontani94} and the extrinsic AHC
\cite{Smit,Bruno} in a Rashba-type 2D electron gas (2DEG) model.
They claimed that the extrinsic-intrinsic crossover occurs
when $\gamma\sim\Delta$, and tried to explain the crossover 
at $\rho\sim1\mu\Omega{\rm cm}$ observed by Asamitsu.
However, it will not be true because $1\mu\Omega{\rm cm}$
corresponds to $\gamma\sim 10^{-3} \approx 8$K,
which is too small for a realistic value of $\Delta$.
(see Fig. \ref{fig:logsxx-12}.)
We stress that Inoue et al. \cite{Inoue} proved that
the intrinsic AHE in a Rashba model almost vanishes due to 
the cancellation by the current vertex 
correction (CVC) by impurities.
On the other hand, we have checked that the CVC for 
anomalous velocity is absent in the present $(d_{xz},d_{yz})$-orbital 
model within the Born approximation, when the impurity potential
is delta-functional.
By this reason, the intrinsic AHC obtained in the present 
work is not modified by the CVC.
We conclude that the present model with atomic $d$-orbital degrees
of freedom is essential for a realistic study of AHE in ferromagnets.

\subsection{Comparison with Experiments: Heavy Fermions}
In paramagnetic heavy fermion (HF) systems,
the Hall coefficients takes huge values due to the AHE,
since the uniform susceptibility $M/B=\chi$ in HF 
is prominently enhanced by the strong Coulomb interaction.
In HF systems, the large quasiparticle damping rate $\hbar/\tau$
comes from the strong electron correlation (or Kondo resonance).
In the early stage,
Coleman et al \cite{Coleman} and Fert and Levy \cite{Fert}
developed theories of the extrinsic AHE:
developed theories of extrinsic AHE:
They studied the extrinsic mechanism based on the $f$-electron impurity 
Anderson models with $d$-orbital channel, and predicted the 
relation $R_{\rm H}\propto \chi \rho$ above $T_0$.
They assumed that the impurity Anderson models is applicable 
for the study of AHE in periodic systems if $T>T_0$.

On the other hand,
Kontani and Yamada found that the $J=5/2$ periodic Anderson
model (PAM) gives a large intrinsic AHC \cite{Kontani94,Kontani97}.
They predict that $\s_{xy}^a \propto \chi$ below $T_0$, 
whereas $\s_{xy}^a \propto \chi\rho^{-2}$ above $T_0$.
We expect this intrinsic AHE is widely observed since
the $J=5/2$ PAM is a well-established effective model 
for Ce-based HF systems:
For example, it could explain a large Van-Vleck susceptibility
in Kondo insulator \cite{Kontani-VV1,Kontani-VV2}.

There remain a lot of future works to be done about the 
AHE in HF systems.
For example, Nakajima et al. found that the AHC in 
two-dimensional HF system, CeMIn$_5$ (M=Co,Rh,Ir)
is negligibly small \cite{Nakajima1,Nakajima2}.
In this compound, strong temperature dependence of $R_{\rm H}$
is given by the normal Hall effect,
due to the effect of the current vertex corrections
(or backflow) in nearly antiferromagnetic Fermi liquids 
 \cite{Kontani99,fromKondo}.
In addition, an interesting field-direction dependence
of AHC is found in CeCu$_6$ or CeCu$_{5.9}$Au$_{0.1}$ \cite{HSato}.
Theoretical studies on these experimental results will give us
important hints to understand the electronic states.

\acknowledgements
We are grateful to J. Inoue, Y. Tanaka, D.S. Hirashima, M. Sato
and H. Sato for enlightening discussions.
We also thank H. Fukuyama for sending us his thesis \cite{Fukuyama}.

\appendix

\section{AHC within the Born approximation;
effect of offdiagonal element of $\gamma$}
 \label{AP-offd}

In previous sections, we studied the AHC by assuming that 
Im$\hat \Sigma$ is diagonal with respect to atomic orbital.
This assumption seems to be justified since 
$|{\rm Im}\Sigma_{1,1}| \gg |{\rm Im}\Sigma_{1,2}|$ when 
$ls$-coupling constant $\l$ is much smaller than bandwidth $W_{\rm band}$.
[When $\l=0$, ${\rm Im}\Sigma_{1,2}=0$ since the local Green 
function ${\hat g}(\w)=\sum_\k {\hat G}_\k(\w)$ is diagonal
due to the orthonormality of $d$-orbitals.]
Here, we derive the AHC within the Born approximation,
by taking both the diagonal and off-diagonal elements of Im$\hat \Sigma$.
As a result, it is confirm that the off-diagonal element is negligible
in calculating the AHC in the present model.

In the Born approximation, the Green function is given by
\begin{widetext}
\begin{eqnarray}
{\hat G}^R(\w)
&=& \frac1{d(\w)}
\left(
\begin{array}{cc}
\w+\mu-\xi_\k^y+i\gamma & \a_\k^0+i\l(1+g-ih) \\
\a_\k^0-i\l(1+g-ih) & \w+\mu-\xi_\k^x+i\gamma
\end{array} 
\right) , \label{eqn:GR2} \\
{\hat G}^A(\w)
&=& \frac1{d(\w)}
\left(
\begin{array}{cc}
\w+\mu-\xi_\k^y-i\gamma & \a_\k^0+i\l(1+g+ih) \\
\a_\k^0-i\l(1+g+ih) & \w+\mu-\xi_\k^x-i\gamma
\end{array} 
\right) \label{eqn:GA2} , 
\end{eqnarray}
\end{widetext}
where $\gamma$, $g$ and $h$ come from the self-energy correction
in the self-inconsistent Born approximation:
\begin{eqnarray}
i\gamma &\equiv& -i{\rm Im}\Sigma_{1,1}^R(0) \nonumber \\
&=&-in_{\rm imp}I^2{\rm Im}\sum_\k \frac{\mu-\xi^y}{d^R(0)}, \label{eqn:Sig1}\\
i\l(g-ih) &\equiv& \Sigma_{1,2}^R(0) \nonumber \\
&=&-i\l n_{\rm imp}I^2 \sum_\k \frac1{d^R(0)}, \label{eqn:Sig2}
\end{eqnarray}
where $n_{\rm imp}$ is the impurity concentration
and $I$ is the impurity potential.
Therefore, 
$\gamma=\pi n_{\rm imp}I^2 N(0)/2 \sim n_{\rm imp}I^2/W_{\rm band}$,
where $N(0)$ is the DOS at the Fermi level.
In the same way, we can estimate that
$g, h \sim n_{\rm imp}(I/W_{\rm band})^2 \sim \g/W_{\rm band} \ll 1$.

We can derive the expression for the AHC by
inserting eqs. (\ref{eqn:GR2}) and (\ref{eqn:GA2})
into the Streda formula in \S \ref{AHE},
in the same way of deriving eqs. (\ref{eqn:sxy-I}) and (\ref{eqn:sxy-II}).
Using the square lattice symmetry of the present model, 
the Fermi surface term at zero temperature is given by
\begin{eqnarray}
\s_{xy}^I&=&
 \frac{2\l}{\pi} \sum_{\k} \frac{v_x^{xx} v_y^{xy}}{d^R(0)d^A(0)}
 \left[ \g(1+g)+h(\mu-\xi^y) \right],
 \nonumber \\
 \label{eqn:sxy-Born}
\end{eqnarray}
Since the main contribution in the $\k$-summation comes from 
$\k\sim \k^*$, where $\k^*$ represents the position of the 
minimum band-splitting $\Delta$, as shown in fig. \ref{fig:FS}.
Since $h|\mu-\xi^y|\sim h\Delta \ll \g$ at $\k=\k^*$,
the second term of eq. (\ref{eqn:sxy-Born}) is negligible.
Therefore, the correction due to off-diagonal element of
the self-energy correction, $i\l(g-ih)$, is very small.
As a result, eq. (\ref{eqn:sxy-I}) is justified
in the Born approximation.

\begin{figure}[htbp]
\includegraphics[width=.99\linewidth]{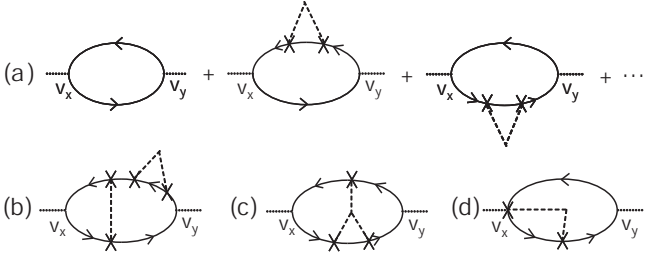}
\caption{
(a) Diagrammatic expression for the intrinsic AHC 
given in eq. (\ref{eqn:sxy-Born}).
Self-energy correction due to impurity scattering
is not diagonal when $\lambda\ne0$.
(b) and (c) represent the Born and the second Born approximation,
respectively.
The latter gives the skew scattering extrinsic AHE
when the impurity potential is non-local.
(d) represents the side jump extrinsic AHC. 
}
  \label{fig:diagram2}
\end{figure}

Figure \ref{fig:diagram2} (a) represents the 
intrinsic AHC given in eq. (\ref{eqn:sxy-Born}).
In good metals, the diagonal self-energy correction does not 
affect the AHC, and the AHC due to off-diagonal self-energy correction
is negligible as we discussed above.
In addition, current vertex correction (CVC) term (b) also exists
in the Born approximation.
Figure (c) and (d) represent other CVC terms
which gives the extrinsic AHC:
(c) gives the skew scattering extrinsic AHC
that is proportional to $\s_{xx}$, and
(d) represents the side-jump extrinsic AHC term
that is proportional to $\{\s_{xx}\}^0$, respectively.
The latter is caused by the lateral displacement of the wave-packet
during the scattering, which is theoretically expressed by the 
modification of velocity operator due to impurity potential
 \cite{Berger,Bruno}.
In the present model, (b)-(d) vanish identically
in the case of local impurity potential.
In contrast, extrinsic AHC occurs
by local impurity potential in 2D electron gas models
when the $ls$-coupling term is $\k$-dependent \cite{Sinitsyn2}.

\section{AHC when $\gamma$ is momentum dependent}
 \label{AP-mom}

In \S \ref{AHE}, \S \ref{Numerical} and \S \ref{BandDiagonal}, 
we have neglected the momentum dependences of the damping rate $\gamma$
to simplify the discussion.
However, this assumption is not realistic because Im$\Sigma_\k(0)$
is usually $\k$-dependent.
In this appendix, we study the AHE by taking the $\k$-dependence of
$\gamma$ into account.
Then, we have to distinguish $\g_x={\rm Im}\Sigma^{xx}(-i0)$ and 
$\g_y={\rm Im}\Sigma^{yy}(-i0)$.
We show that the expression for $\s_{xy}^I$
in eq. (\ref{eqn:ap-xy-I}) could be changed prominently
when the $\k$-dependence of $\g$ is taken into account.

Along with the derivation of eq. (\ref{eqn:sxy-I}),
we obtain the expression for $\s_{xy}^I$ in the present case:
\begin{eqnarray}
\s_{xy}^I&=&
\frac{\l }{\pi}\sum_{\k} 
\frac{v_x^{xx} v_y^{xy}\gamma_y + v_x^{xy} v_y^{yy}\gamma_x}{d^R(0)d^A(0)} ,
 \label{eqn:B-I}
\end{eqnarray}
where
\begin{eqnarray}
d^R(0)d^A(0)=((\mu-E_\k^+)^2+\g_+^2)((\mu-E_\k^-)^2+\g_-^2) .
\end{eqnarray}
When both $\g_x$ and $\g_y$ are very small, 
\begin{eqnarray}
\g^\pm &=& \frac{\g_x+\g_y}{2} 
\left( 1\pm \frac{\xi^x-\xi^y}{E_\k^+-E_\k^-}
\frac{\g_x-\g_y}{\g_x+\g_y} \right) .
\end{eqnarray}
In this case, eq. (\ref{eqn:B-I}) is given by
\begin{eqnarray}
\s_{xy}^I&=&
 \l \sum_{\k} 
\frac{v_x^{xx}v_y^{xy}\gamma_y + v_x^{xy} v_y^{yy}\gamma_x}{(E_\k^+-E_\k^-)^2}
 \nonumber \\
& &\times \left[\delta(\mu-E_\k^+)\frac1{\gamma_\k^+}
 + \delta(\mu-E_\k^-)\frac1{\gamma_\k^-} \right] 
 \label{eqn:B-IG}
\end{eqnarray}
When $\g_\mu$ is momentum dependent,
$\g_{x(y)}/\gamma^+, \g_{x(y)}/\gamma^- \ne 1$ except for $|k_x|=|k_y|$.
Therefore, eq. (\ref{eqn:B-IG}) is not equal to eq. (\ref{eqn:ap-xy-I})
even in good metals ($\gamma_x, \gamma_y \rightarrow 0$).

In general, the momentum dependence (and the band-dependence)
of $\gamma$ could give rise to a nontrivial modulation of
the AHC due to the Fermi surface term, $\s_{xy}^I$.
This effect will be prominent in strongly correlated electron systems
like high-$T_{\rm c}$ cuprates and heavy Fermion systems.
On the other hand, 
the Fermi sea term $\s_{xy}^{II}$, which is 
given by eqs. (\ref{eqn:ap-xy-II1}) and (\ref{eqn:ap-xy-II2}), 
is satisfied when $\gamma_x, \gamma_y \rightarrow 0$,
even if $\g_x/\g_y\ne 1$.

\section{Calculation of AHC
when $\gamma$ is band dependent}
 \label{AP-band}

As shown in Appendix \ref{AP-mom},
we have shown that the expression for Fermi surface term 
$\s_{xy}^I$ given in eq. (\ref{eqn:ap-xy-I}) is modified 
by the $\k$-dependence of $\g$.
In this section, we explain that eq. (\ref{eqn:ap-xy-I})
is also changed when $\g$ depends on bands.
This fact was already pointed out in Appendix D
of ref. \cite{Kontani94}.
Here, we derive the first order term of $\s_{xy}^I$
with respect to the $ls$-coupling, and correct a mistake
in Appendix D of ref. \cite{Kontani94}.
By extracting a single $ls$-coupling term from eq. (\ref{eqn:AP-I}),
we obtain
\begin{eqnarray}
\s_{xy}^I&=&
 \frac{\l }{2\pi}\sum_{\k,l\ne m} G_l^RG_l^A 
\nonumber \\
& &\times \left[ v_x^{l l}( G_m^R v_y^{m l} \langle l|l_z|m\rangle 
 + G_m^A v_y^{l m} \langle m|l_z|l\rangle ) \right.
 \nonumber \\
& &\left. + (v_x^{l m} G_m^R \langle m|l_z|l\rangle
 + v_x^{m l} G_m^A \langle l|l_z|m\rangle ) v_y^{l l} \right] ,
\end{eqnarray}
where $G_l^R=(\mu-E_\k^l +i\gamma_l)^{-1}$ and $v_\mu^{l m}$ are 
the Green function and the velocity without $ls$-coupling term.
The diagrammatic expression is given in fig. \ref{fig:diagram} (b).
Using the relation 
$\langle l|l_z|m\rangle = -\langle m|l_z|l\rangle$
and $v_\mu^{l m}= v_\mu^{m l}$, we obtain that
\begin{eqnarray}
\s_{xy}^I&=&
 \frac{\l }{2\pi}\sum_{\k,l\ne m} G_l^RG_l^A 
(G_m^R- G_m^A) \langle l|l_z|m\rangle
 \nonumber \\
& &\times \left[ v_x^{l l}v_y^{l m} - v_x^{l m}v_y^{l l} \right] .
 \label{eqn:ApC-I}
\end{eqnarray}
When $\gamma_l$ is very small, 
\begin{eqnarray}
& &G_l^RG_l^A(G_m^R-G_m^A) 
= \frac{2\pi i}{(E_\k^l-E_\k^m)^2}
 \nonumber \\
& &\ \ \ \ \ \ \ \ \ \ \times
\left[ \delta(\mu-E_\k^l) \frac{\g_m}{\g_l}
 + \delta(\mu-E_\k^m) \right] .
\end{eqnarray}
In this case, eq. (\ref{eqn:ApC-I}) is given by
\begin{eqnarray}
\s_{xy}^I&=&
 \l  \sum_{\k,l\ne m} 
\left[ \delta(\mu-E_\k^l) \frac{\g_m}{\g_l} + \delta(\mu-E_\k^m) \right]
 \nonumber \\
& &\times
 \frac{i\langle l|l_z|m\rangle}{(E_\k^l-E_\k^m)^2}
 \left[ v_x^{l l}v_y^{l m} - v_x^{l m}v_y^{l l} \right] .
 \label{eqn:ApC-I2}
\end{eqnarray}

In summary, $\s_{xy}^I$ depends on the ratio of $\g_m/\g_l$.
$\s_{xy}^I$ could take a huge value if $\g_l$ differs much
from band to band.
This fact was already pointed out by ref. \cite{Kontani94}.
On the other hand, 
eqs. (\ref{eqn:ap-xy-II1}) and (\ref{eqn:ap-xy-II2}) for
the Fermi sea term $\s_{xy}^{II}$
is satisfied when $\g_l \rightarrow 0$,
even if $\g_m/\g_l\ne 1$.
The present result will be valid even if the $ls$-coupling 
is treated unperturbatively, as done in \S \ref{Model} - \S \ref{Numerical}. 
In fact, Fig. \ref{fig:lambda} shows that $\s_{xy}$ is approximately
linear-in-$\lambda$ for $\lambda<0.2$.

In Appendix D of ref. \cite{Kontani94}, $E_\k^m$ in 
$\delta(\mu-E_\k^m)$ was replaced with $E_\k^l$ by mistake.
Thus, eq. (\ref{eqn:ApC-I2}) is not equal to the KL's term
which is given by [eq. (3.5)]$\times (2/m E_b)$ in ref. \cite{KL} 
with $(a,b)=(x,y)$. 
This difference is natural since the former and the latter are
the Fermi surface term and the Fermi sea term, respectively.



\begin{thebibliography}{99}

\bibitem{Yoshii}
 S. Yoshi, S. Iikubo, T. Kageyama, K. Oda, Y. Kondo, K. Murata 
 and M. Sato: J. Phys. Soc. Jpn. {\bf 69} (2000) 3777.

\bibitem{Yasui}
 Y. Yasui, T. Kageyama, T. Moyoshi, M. Soda, M. Sato and K. Kakurai:
 J. Phys. Soc. Jpn. {\bf 75} (2006) 084711.

\bibitem{Taguchi1}
 Y. Taguchi and Y. Tokura: Phys. Rev. B {\bf 60} (1999) 10280.

\bibitem{Taguchi2}
 Y. Taguchi, Y. Oohara, H. Yoshizawa, N. Nagaosa and Y. Tokura: 
 Science {\bf 291} (2001) 2573.

\bibitem{Ong}
W.L. Lee, S. Watauchi, V.L. Miller, R.J. Cava and N.P. Ong:
Science {\bf 303} 1647.

\bibitem{Reiner}
 L. Klein, J.R. Reiner, T.H. Geballe, M.R. Beasley, and A. Kapitulnik:
 Phys. Rev. B {\bf 61} (2000) R7842.

\bibitem{Izumi}
 M. Izumi, K. Nakazawa, Y. Bando, Y. Yoneda and H. Terauchi:
 J. Phys. Soc. Jpn. {\bf 66} (1997) 3893.

\bibitem{HSato2}
 H. Sato, T. Kumano, Y. Aoki, T. Kaneko and R. Yamamoto:
 J. Phys. Soc. Jpn. {\bf 62} (1993) 416.

\bibitem{Canedy}
 C.L. Canedy, X.W. Li and G. Xiao: Phys. Rev. B {\bf 62} (2000) 508.

\bibitem{Tatara}
 G. Tatara and H. Kawamura: J. Phys. Soc. Jpn. {\bf 71} (2002) 2613.

\bibitem{Kawamura}
 H. Kawamura: Phys. Rev. Lett. {\bf 90} (2003) 047202.

\bibitem{KL}
 R. Karplus and J. M. Luttinger:  Phys. Rev. {\bf 95} (1954) 1154.

\bibitem{Smit}
 J. Smit: Physica {\bf 24} (1958) 39.

\bibitem{KL2}
 J. M. Luttinger: 
 Phys. Rev. {\bf 112} (1958) 739.

\bibitem{Fukuyama}
H. Fukuyama, Ph. D thesis, 1970 (unpublished). 

\bibitem{Sinitsyn2}
 N.A. Sinitsyn, A.H. MacDonald, T. Jungwirth, V.K. Dugaev and J. Sinova:
 cond-mat/0608682.

\bibitem{Berger}
 L. Berger, Phys. Rev. B {\bf 2} (1970) 4559.

\bibitem{Asamitsu}
 T. Miyasato, N. Abe, T. Fujii, A. Asamitsu, S. Onoda, 
 Y. Onose, N. Nagaosa, Y. Tokura: cond-mat/0610324.

\bibitem{Bruno}
 A. Crepieux and P. Bruno: Phys. Rev. B {\bf 64} (2001) 014416.

\bibitem{Streda}
 P. Streda: J. Phys. C: Solid State Phys. {\bf 15} (1982) L717.

\bibitem{Mac}
 N.A. Sinitsyn, J.E. Hill, H. Min, J. Sinova and A.H. MacDonald:
 Phys. Rev. Lett. {\bf 97} (2006) 106804.

\bibitem{Murakami}
 S. Murakami, N. Nagaosa and S.C. Zhang: Science {\bf 301} (2003) 1348.

\bibitem{future}
 H. Kontani, T. Tanaka, D.S. Hirashima, K. Yamada and J. Inoue, 
 cond-mat/cond-mat/0702447.

\bibitem{MOnoda}
 M. Onoda and N. Nogaosa: J. Phys. Soc. Jpn. {\bf 71} (2002) 19.

\bibitem{Niu}
 G. Sundaram and Q. Niu: Phys. Rev. B {\bf 59} (1999) 14915.

\bibitem{Kontani94}
 H. Kontani and K. Yamada:
 J. Phys. Soc. Jpn. {\bf 63} (1994) 2627.
 
\bibitem{Kontani97}
 H. Kontani and K. Yamada:
 J. Phys. Soc. Jpn. {\bf 66} (1997) 2252.

\bibitem{Onuki1}
Y. \=Onuki, T. Yamayoshi, I. Ukon, T. Komatsubara, 
A. Umezawa, W. K. Kwok, G. W. Crabtree and D. G. Hinks: 
J. Phys. Soc. Jpn. {\bf 58} (1989) 2119.

\bibitem{Onuki2}
Y. \=Onuki, T. Yamayoshi, T. Omi, I. Ukon, A. Kobori 
and T. Komatsubara: J. Phys. Soc. Jpn. {\bf 58} (1989) 2126.

\bibitem{Miyazawa}
 M. Miyazawa, H. Kontani and K. Yamada:
 J. Phys. Soc. Jpn. {\bf 68} (1999) 1625.

\bibitem{Fe}
 Y. Yao, L. Kleinman, A.H. MacDonald, J. Sinova, T. Jungwirth,
 D.S. Wang, E. Wang and Q. Niu: Phys. Rev. Lett. {\bf 92} (2004) 037204.

\bibitem{Fang}
 Z. Fang et al.: Science {\bf 302} (2003) 92. 

\bibitem{Sigrist}
K.K. Ng and M. Sigrist, Europhys. Lett. {\bf 49}, 473 (2000).

\bibitem{Nomura}
 T. Nomura and K. Yamada: J. Phys. Soc. Jpn. {\bf 71} (2002) 404; 
 T. Nomura and K. Yamada: J. Phys. Soc. Jpn. {\bf 71} (2002) 1993; 

\bibitem{Yanase}
Y. Yanase and M. Ogata, J. Phys. Soc. Jpn. {\bf 72}, 673 (2003).

\bibitem{Sinitsyn1}
 N.A. Sinitsyn, Q. Niu, J. Sinova and K. Nomura:
 cond-mat/0502426.

\bibitem{Kondo}
 J. Kondo: Prog. Theor. Phys. {\bf 27} (1962) 772.

\bibitem{Coleman}
 P. Coleman, P. W. Anderson and T. V. Ramakrishnan: 
 Phy. Rev. Lett. {\bf 55} (1985) 414.

\bibitem{Fert}
 A. Fert and P. M. Levy: Phy. Rev. B {\bf 36} (1987) 1907.

\bibitem{Slater}
 J. Slater and G. Koster, Phys. Rev. {\bf 94}, 1498 (1954). 

\bibitem{Nakano}
 H. Nakano: Int. J. Mod. Phys. B {\bf 7} (1993) 2397.

\bibitem{Thouless}
 D.J. Thouless, M. Kohmoto, M.P. Nightingale and M. den Nijs:
 Phys. Rev. Lett. {\bf 49} (1982) 405.

\bibitem{AHE-CSRO}
J. Jin et al., cond-mat/0112405.

\bibitem{Haldane}
 F.D.M. Haldane: Phys. Rev. Lett. {\bf 61} (1998) 2015.

\bibitem{SOnoda}
 S. Onoda, N. Sugimoto and N. Nagaosa: 
 Phys. Rev. Lett. {\bf 97} (2006) 126602.

\bibitem{Inoue}
J. Inoue, T. Kato, Y. Ishikawa, H. Itoh, G.E.W. Bauer and  L.W. Molenkamp:
 Phys. Rev. Lett. {\bf 97} (2006) 046604.

\bibitem{Kontani-VV1} 
 H. Kontani and K. Yamada:
 J. Phys. Soc. Jpn. {\bf 65} (1996) 172.

\bibitem{Kontani-VV2} 
 H. Kontani and K. Yamada:
 J. Phys. Soc. Jpn. {\bf 66} (1997) 2232.

\bibitem{Nakajima1}
 Y. Nakajima, K. Izawa, Y. Matsuda, S. Uji, T. Terashima, H. Shishido, 
 R. Settai, Y. Onuki and H. Kontani, J. Phy. Soc. Jpn. {\bf 73} (2004) 5.

\bibitem{Nakajima2}
 Y. Nakajima, K. Izawa, Y. Matsuda, K. Behnia, H. Kontani, M. Hedo, 
 Y. Uwatoko, T. Matsumoto, H. Shishido, R. Settai and Y. Onuki:
  J. Phy. Soc. Jpn. {\bf 75} (2006) 023705.

\bibitem{Kontani99}
 H. Kontani, K. Kanki and K. Ueda: Phys. Rev. B {\bf 59} (1999) 14723.

\bibitem{fromKondo}
 H. Kontani and K. Yamada: J. Phy. Soc. Jpn. {\bf 74} (2005) 155.

\bibitem{HSato}
 T. Namiki, H. Sato, H. Sugawara, Y. Aoki, R. Settai and Y. Onuki:
 preprint.



\end{thebibliography}
\end{document}